\documentclass[traditabstract]{aa} 
\usepackage{graphicx}
\usepackage{natbib}
\usepackage{bm}
\usepackage{txfonts}
\begin{document}
   \title{Nonlinear energy transfers in accretion discs MRI turbulence\\I-Net vertical field case}
   
   \author{G. Lesur\inst{1,2} and P.-Y. Longaretti\inst{2} }
   \institute{Department of Applied Mathematics and Theoretical Physics, University of Cambridge, Centre for Mathematical Sciences,
Wilberforce Road, Cambridge CB3 0WA, UK\\  \email{geoffroy.lesur@obs.ujf-grenoble.fr}  \and Laboratoire d'Astrophysique, UJF CNRS, BP 53, 38041 Grenoble Cedex 9, France 
                   }

   \date{Received date / Accepted date}

 
  \abstract{The magnetorotational instability (MRI) is believed to be responsible for most of the angular momentum transport in accretion discs. However, molecular dissipation processes may drastically change the efficiency of MRI turbulence in realistic astrophysical situations. The physical origin of this dependency is still poorly understood as linear and quasi linear theories fail to explain it. In this paper, we look for the link between molecular dissipation processes and MRI transport of angular momentum in non stratified shearing box simulations including a mean vertical field. We show that magnetic helicity is unimportant in the model we consider. We perform a spectral analysis on the simulations tracking energy exchanges in spectral space when turbulence is fully developed. We find that the energy exchanges are essentially direct (from large to small scale) whereas some non linear interactions appear to be non local in spectral space. We speculate that these non local interactions are responsible for the correlation between turbulent transport and molecular dissipation. We argue that this correlation should then disappear when a significant scale separation is achieved and we discuss several methods by which one can test this hypothesis.}


   \titlerunning{MRI nonlinear energy transfers}
   \maketitle
%

\section{Introduction}
The transport of angular momentum in astrophysical discs is a central problem of accretion theory. To explain discs lifetime and accretion rates, it is often assumed that these objects are turbulent. Turbulence is then included in global models using a turbulent viscosity prescription as in the $\alpha$ disc model \citep{SS73}. 

The origin of this turbulence has been the subject of many debates over the past decades. It is now generally assumed that the magnetorotational instability, or shortly MRI \citep{V59,C60,BH91a}, is responsible for disc turbulence, although hydrodynamic processes might also be at work \citep{LO10,LP10}. Although MRI generated turbulence is generally efficient at transporting angular momentum \citep{HGB95}, recent results have shown a strong sensitivity of MRI turbulence on small scale dissipation processes \citep{LL07,FPLH07}, and in particular on the magnetic Prandtl number $Pm$ (ratio of microscopic viscosity to resistivity). This effect, called the $\alpha-Pm$ correlation, casts doubts on the actual efficiency of the MRI in realistic situations since $Pm$ can vary by several order of magnitude in discs \citep{BH08}. Several attempts have been made to explain this correlation, either from the linear theory of dissipative MRI modes \citep{PC08} or from the quasi-linear parasitic modes theory \citep{PG09}. However, these approaches were shown to be unsuccessful when compared to high Reynolds number simulations \citep{LL10}. Instead, \cite{LL10} suggested that the $\alpha-Pm$ correlation could be due to the nature of the MHD cascade in MRI generated turbulence, in which one might expect inverse cascades and/or non local interaction in spectral space. This kind of process would allow for a direct communication between the injection scales (transport scales) and the largest dissipation scale (either resistive or viscous).

The purpose of the present work is to investigate some of the conjectures presented by \cite{LL10} regarding the nature of the MHD turbulent cascade in accretion discs. To this end, we consider several of the high resolution simulations presented by \cite{LL10} and we analyse the energy exchanges in spectral space. This paper is organized as follows. We describe our model, equations and the spectral analysis we use in section \ref{equations}. Section \ref{results} is the core of this paper and discusses our numerical results. The implications of these results are presented in the final section. 
\section{Model and spectral analysis\label{equations}}

\subsection{Equations}
In the following, we will adopt the shearing box model which accurately represent the local physics of an accretion disc \citep{HGB95,B03,RU08}. The adopted coordinate system is such that $x=(r-r_0)$ and $y=r_0\phi$ where $r_0$ is the fiducial radius of the shearing box in the disc, and $\phi$ the azimuthal coordinate in the rotating frame. The velocity can be decomposed as a mean velocity plus a fluctuating part $\bm{U}= -q\Omega x \bm{e_y}  + \bm{v}$ where $\Omega$ is the local rotation rate and $q=3/2$ for a Keplerian disc rotation profile. As a simplification, we assume the flow is incompressible, which corresponds to the ``small shearing box limit'' of \cite{UR04}. The equations of motion then read
\begin{eqnarray}
\nonumber \partial _t \bm{v}&=& -\bm{v}\cdot\bm{\nabla}\bm{v}-\bm{\nabla} \Pi +
\bm{B}\cdot\bm{\nabla}\bm{B}+q\Omega x\partial_y \bm{v}\\
\label{motion}& & +2\Omega v_y\bm{e_x}-(2-q)\Omega v_x \bm{e_y}+\nu\bm{\nabla}^2\bm{ v},\\
 \nonumber \partial _t \bm{B}&=&-\bm{v}\cdot\bm{\nabla}\bm{B}+\bm{B}\cdot\bm{\nabla}\bm{v}\\
 \label{induction}& & +q\Omega x\partial_y \bm{B}-q\Omega B_x\bm{e_y}+\eta \bm{\nabla}^2\bm{ B},\\
\label{Vstruct} \bm{\nabla \cdot v}&=&0,\nonumber\\
\label{Bstruct}\bm{\nabla \cdot B}&=&0,\nonumber
\end{eqnarray}

\noindent where $\Pi$ is the total pressure, $\nu$ the viscosity and $\eta$ the ohmic resistivity. In the following, we impose a mean vertical field $B_0$ which will be conserved during the evolution of the flow thanks to the shearing-sheet boundary conditions. It should be noted that the magnetic field strength is expressed in Alfv\'en speed for simplicity.

Several dimensionless numbers characterise the equations of motions. In this paper, we will use the following:
\begin{itemize}
\item The amplitude of the imposed mean vertical field measured by
\begin{equation}
\beta = \frac{(q\Omega)^2 L_z^2}{B_0^2}
\end{equation}
where $L_z$ is the vertical box size. This definition mimics the usual plasma $\beta$ in vertically stratified discs obeying the vertical hydrostatic equilibrium constraint $c_s\sim \Omega L_z$
\item The viscous Elsasser number:
\begin{equation}
\Lambda_\nu=\frac{B_0^2}{\Omega\nu}
\end{equation}
which is related to the Reynolds number used in \cite{LL10} by $\Lambda_\nu=qRe/\beta$
\item The resistive Elsasser number:
\begin{equation}
\Lambda_\eta=\frac{B_0^2}{\Omega \eta}
\end{equation}
connected to the magnetic Reynolds number by a similar relation
\item The magnetic Prandtl number
\begin{equation}
Pm=\frac{\nu}{\eta}=\frac{\Lambda_\eta}{\Lambda_\nu}
\end{equation}
which compares the amount of viscous and resistive dissipation.
\end{itemize}

\subsection{Fourier transform in sheared flows}
It is convenient to introduce the shearing frame ($x'$, $y'$, $z'$) :
\begin{eqnarray}
\nonumber x &=& x',\nonumber\\
\nonumber y &=& y'-q\Omega t x',\nonumber\\
\nonumber z &=& z'.\nonumber
\end{eqnarray}
Writing the equations of motions in the sheared frame allows us to eliminate the explicit spatial dependency:
\begin{eqnarray}
\nonumber \partial _t \bm{v}&=& -\bm{v}\cdot\bm{\nabla}\bm{v}-\bm{\nabla} \Pi +
\bm{B}\cdot\bm{\nabla}\bm{B}\\
\label{motion-shear}& & +2\Omega v_y\bm{e_x}-(2-q)\Omega v_x \bm{e_y}+\nu\bm{\nabla}^2\bm{ v},\\
 \nonumber \partial _t \bm{B}&=&-\bm{v}\cdot\bm{\nabla}\bm{B}+\bm{B}\cdot\bm{\nabla}\bm{v}\\
 \label{induction-shear}& & -q\Omega B_x\bm{e_y}+\eta \bm{\nabla}^2\bm{ B}.
\end{eqnarray}
where $\bm{v}$ and $\bm{B}$ are now assumed to be functions of $\bm{x'}$ so that the nabla operator expression becomes 
\begin{equation}
\label{nablaOp}
\nonumber \bm{\nabla}=\bm{e_x}(\partial_{x'}+q\Omega t \partial_{y'})+\bm{e_y}\partial_{y'}+\bm{e_z}\partial_{z'}
\end{equation}

In the sheared frame, the shearing sheet boundary conditions make every physical quantity $X(\bm{x}')$ periodic so that $X$ can be expanded in Fourier series:
\begin{equation}\label{fourier}
\label{SpectrumDef}
X(\bm{x}')=\sum_{\bm{k}'} X_{\bm{k}'} \exp\left(i {\bm{k}'}\cdot {\bm{x}'}\right) = \sum_{\bm{k(t)}} X_{\bm{k(t)}} \exp\left(i {\bm{k}(t)}\cdot {\bm{x}}\right).
\end{equation}
This relation defines the \emph{time dependent} unsheared wave vectors:
\begin{eqnarray}
\label{shwaveX} k_x &=& k_{x'}+q\Omega k_{y'} t,\\
\label{shwaveY} k_y &=& k_{y'},\\
\label{shwaveZ} k_z &=& k_{z'}.
\end{eqnarray}
As expected, the application of the $\bm{\nabla}$ operator to $X(\bm{x}')$ corresponds to a multiplication of its Fourier components by $i\bm{k}(t)$. 

This definition of the unsheared wave vectors with the Fourier decomposition (\ref{SpectrumDef}) is usually referred to as a ``shearing wave'' decomposition of a sheared flow. It was first used by Lord Kelvin to study the stability of sheared flows \citep{T87} and later in the astrophysical context by \cite{GL65} for spiral arms of galaxies.

\subsection{Shell filter decomposition}

Following \cite{FR95} and \cite{AMP07}, one defines shell filtered quantities in the unsheared Fourier space. At any given time, a series of linearly spaced shell sizes $K$ is defined from $K_1$ to $K_{\mathrm{max}}$; by construction $\delta K = K_j-K_{j-1}$ is the shell width (for any $j$). We define the shell-filtered field $X_{K_j}$ in shell $K_j$ by:
\begin{equation}\label{Xshell}
X_{K_j} = \sum_{K_j-\delta K/2 < \bm{k} \le K_j+\delta K/2} X_{\bm{k}} \exp\left(i {\bm{k}}\cdot {\bm{x}}\right).\nonumber
\end{equation}
It should be noted that since $\bm{k}$ depends on time, the exact number and the distribution of the modes entering the above formula for any given $K$ might vary in time, adding an extra complication compared to the homogeneous case of \cite{AMP07}. This point is discussed in the Appendices.

\subsection{Energy transfer equations}

The transfers we are interested in relate to the equations of the kinetic and magnetic energies. These shell-restricted, box-averaged equations involve a number of transfer functions that are introduced along with the related equations. We follow here the logic of \cite{AMP07} and extend it to shear flows. The analysis of transfers gives indications about the locality of interactions in Fourier space.

Because of the incompressibility condition, energy transfers (but also stress and magnetic helicity) involve only at most the product of three components of the velocity and magnetic fields and their derivatives. Therefore, couplings in Fourier space depend only on triads of wave-vectors, noted $\bm{k_1}$, $\bm{k_2}$ $\bm{k_3}$. The closing condition ($\bm{k_1}+\bm{k_2}=\bm{k_3}$) furthermore imposes that at least two of the wave vectors $\bm{k_i}$ are of the same magnitude; the third one can be either much smaller (implying nonlocal  couplings through large scales) or of the same magnitude as the other two (implying locality of couplings in Fourier space).

Using the equations of motion in the sheared frame (\ref{motion-shear}---\ref{induction-shear}), one can derive the equation for the shell filtered energy density:
\begin{eqnarray}
\nonumber \partial_t E_K &=& \sum_Q \left[T_{vv}(Q,K)+T_{bv}(Q,K)\right]+S_{v,K}\\
\label{EqEk}& &+q\Omega I_{v,K} -\nu D_{v,K}\\
\nonumber \partial_t M_K &=& \sum_Q \left[T_{bb}(Q,K)+T_{vb}(Q,K)\right]+S_{b,K}\\
\label{EqEm}& &+ q\Omega I_{b,K}-\eta D_{b,K}
\end{eqnarray}
\noindent where $E_K=\langle\bm{v}^2_K/2\rangle$ and $M_K=\langle\bm{B}^2_K/2\rangle$. In the above expression, we have defined the transfer functions by

\begin{eqnarray}
T_{vv}(Q,K) & = & -\langle\bm{v}_K\cdot (\bm{v}\cdot\bm{\nabla})\bm{v}_Q\rangle ,\nonumber\\
T_{bb}(Q,K) & = & -\langle\bm{B}_K\cdot (\bm{v}\cdot\bm{\nabla})\bm{B}_Q\rangle ,\nonumber\\
T_{bv}(Q,K) & = & \langle \bm{v}_K\cdot (\bm{B} \cdot\bm{\nabla})\bm{B}_Q\rangle ,\nonumber\\
T_{vb}(Q,K) & = &  \langle\bm{B}_K\cdot (\bm{B} \cdot\bm{\nabla})\bm{v}_Q\rangle ,\nonumber
\end{eqnarray}

\noindent where $\langle\ .\ \rangle$ denotes an spatial average on the shearing box volume. $T_{ij}(Q,K)$ represents the transfer from energy `$i$' (kinetic or magnetic) from shell $Q$ to energy `$j$' (kinetic or magnetic) in shell $K$. Note that $T_{ij}(Q,K)=-T_{ji}(K,Q)$, so that whatever is taken from one shell of one type is totally transferred to the other shell. Similarly, $T_{ii}(K,K)= 0$: there is no effective transfer from a shell to itself, as should be. This justifies the identification of these quantities as shell-to-shell energy transfer functions; in effect, the third member of a triad is only a relay in effective energy exchanges between shells $Q$ and $K$ (for a more detailed discussion, see \citealt{V04}). In these expressions, $\bm{v}_K\cdot\bm{v}_K=\bm{v}_K\cdot\bm{v}$, has been used, as well as $\bm{v}=\sum_Q\bm{v}_Q$.

The next terms in the shell energy budget involve energy transfers due to the mean shear $S_{v,K}$ and $S_{b,K}$. These terms are singular in time as they correspond to energy fluctuations due to wave entering or leaving the shell $K$ as $k(t)$ evolves. They can be formally defined by:
\begin{eqnarray}
\nonumber S_{X,K}&=&\sum_{k'}\frac{X_{\bm{k'}}^* X_{\bm{k'}}}{2}\delta(t-t_{k'}) \epsilon_{k'}\\
\nonumber &=& \sum_{k'}     \frac{q\Omega k_y k_x(t)}{|k(t)|}\frac{X_{\bm{k'}}^* X_{\bm{k'}}}{2}\\
\nonumber & &\times \Big[\delta\Big(|k(t)|-K+\delta K/2\Big)- \delta\Big(-|k(t)|+K+\delta K/2\Big)\Big].
\end{eqnarray}
where $t_{k'}$ is the instant when the wave $\bm{k'}$ enters or exits the shell $K$ and $\epsilon_{k'}=\pm 1$ for an entering/exiting wave (see appendix \ref{shearTransfer}-\ref{shearTransfer2}).

The remaining terms in the shell kinetic and magnetic energy budgets, 
\begin{eqnarray}
 I_{v,K} & = &  \langle v_{x,K}v_{y,K} \rangle,\nonumber\\
 I_{b,K} & = & - \langle B_{x,K}B_{y,K}\rangle ,\nonumber\\
 D_{v,K} & = &  \langle (\bm{\nabla} \times \bm{v}_K)^2\rangle,\nonumber\\
 D_{b,K} & = &  \langle (\bm{\nabla} \times \bm{B}_K)^2\rangle,\nonumber
\end{eqnarray}

\noindent represent the energy injection through the shear and dissipation through viscosity and resistivity in shell $K$. 

\subsection{Energy fluxes in spectral space}

Using the transfer function defined above, it is possible to introduce energy flux in Fourier space. In this work, we will use the following fluxes:
\begin{eqnarray}
\label{Eq:Flux1}
\mathcal{F}_{v}(K_0,t)&=&\sum_{K=K_0}^{K_{\mathrm{max}}}\sum_Q T_{vv}(Q,K)\\
\mathcal{F}_{b}(K_0,t)&=&\sum_{K=K_0}^{K_{\mathrm{max}}}\sum_Q T_{bb}(Q,K)\\
\mathcal{F}_{x}(K_0,t)&=&\sum_{K=K_0}^{K_{\mathrm{max}}}\sum_Q T_{vb}(Q,K)+T_{bv}(Q,K)\\
\mathcal{F}_{s,v}(K,t)&=&\sum_{k'}  \frac{q\Omega k_y k_x(t)}{|k(t)|}\frac{\bm{V}_{\bm{k'}}^*\cdot \bm{V}_{\bm{k'}}}{2}\delta\big(|k(t)|-K\big)\\
\label{Eq:Flux2}
\mathcal{F}_{s,b}(K,t)&=&\sum_{k'}  \frac{q\Omega k_y k_x(t)}{|k(t)|}\frac{\bm{B}_{\bm{k'}}^*\cdot \bm{B}_{\bm{k'}}}{2}\delta\big(|k(t)|-K\big)
\end{eqnarray}
where $\mathcal{F}_v$, $\mathcal{F}_b$, $\mathcal{F}_x$ and $\mathcal{F}_s$ are respectively the kinetic, magnetic, exchange and shear fluxes; the shear fluxes are evaluated in $K=K_0+\delta K/2$. Each of these fluxes computes the amount the energy transferred from shells $K\le K_0$ to shells $K>K_0$ (i.e. the flux of energy ``through'' the outer boundary of shell $K_0$). The kinetic (respectively magnetic) fluxes compute transfers of kinetic to kinetic (respectively magnetic to magnetic) energy, whereas the exchange flux is a flux of \emph{total} energy (magnetic plus kinetic) in which magnetic and kinetic energy are constantly transformed into one another. Shear fluxes are singular in time as they are non zero only when a wave enters or leave shells $K>K_0$. It should be noted that shear fluxes are statistically non zero only for anisotropic turbulence, as the amplitude of any mode $(k_x,k_y,k_z)$ should be statistically equal to the amplitude of the mode $(-k_x,k_y,k_z)$ in isotropic turbulence.

These fluxes allow one to check the direction of the energy transfers due to the nonlinear terms. Indeed, a direct energy transfer (large to small scale) implies a positive flux with the above definition, whereas an inverse cascade can be characterised by a negative flux. 

 \begin{figure*}[t]
   \centering
   \includegraphics[width=0.45\linewidth]{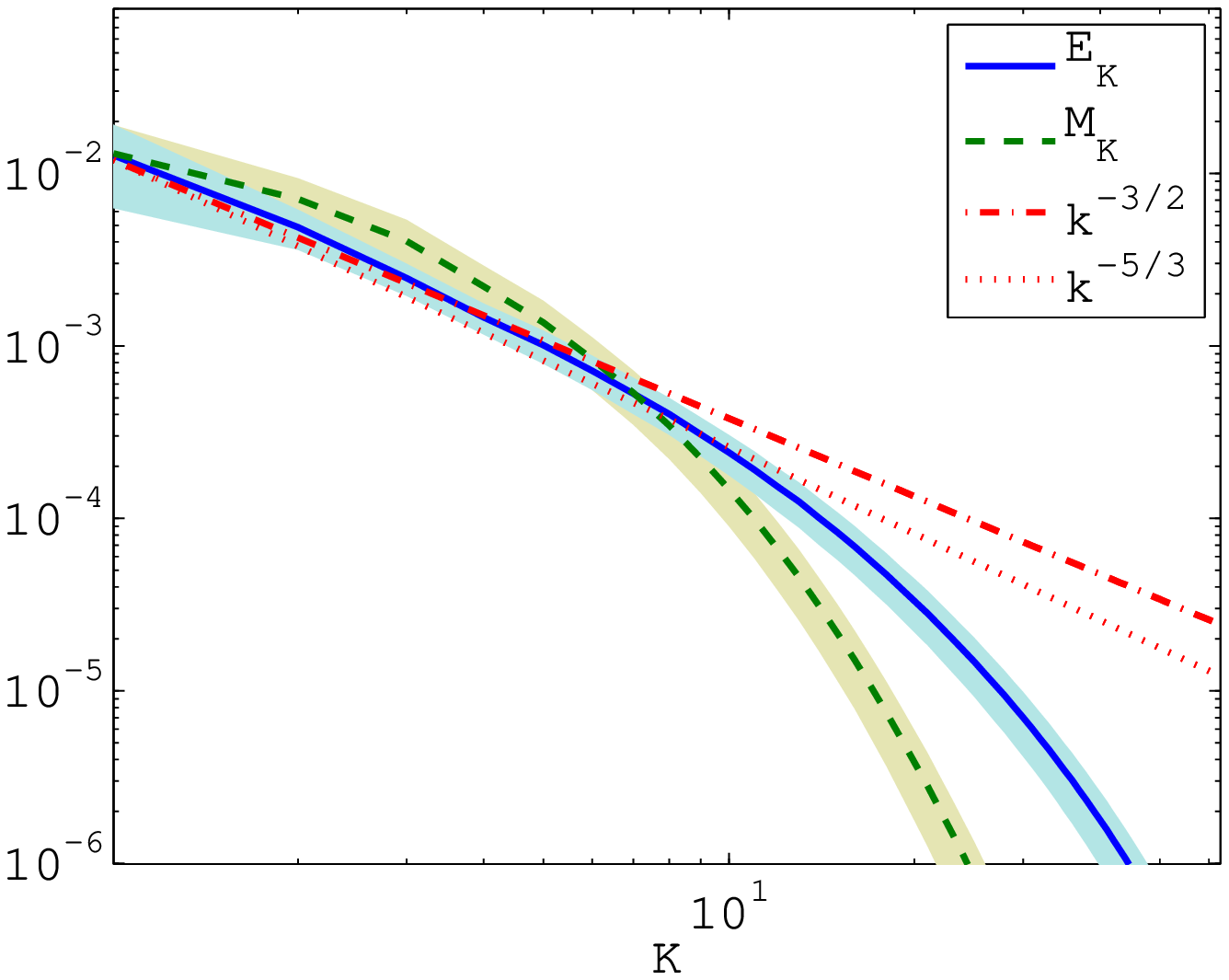}
   \includegraphics[width=0.45\linewidth]{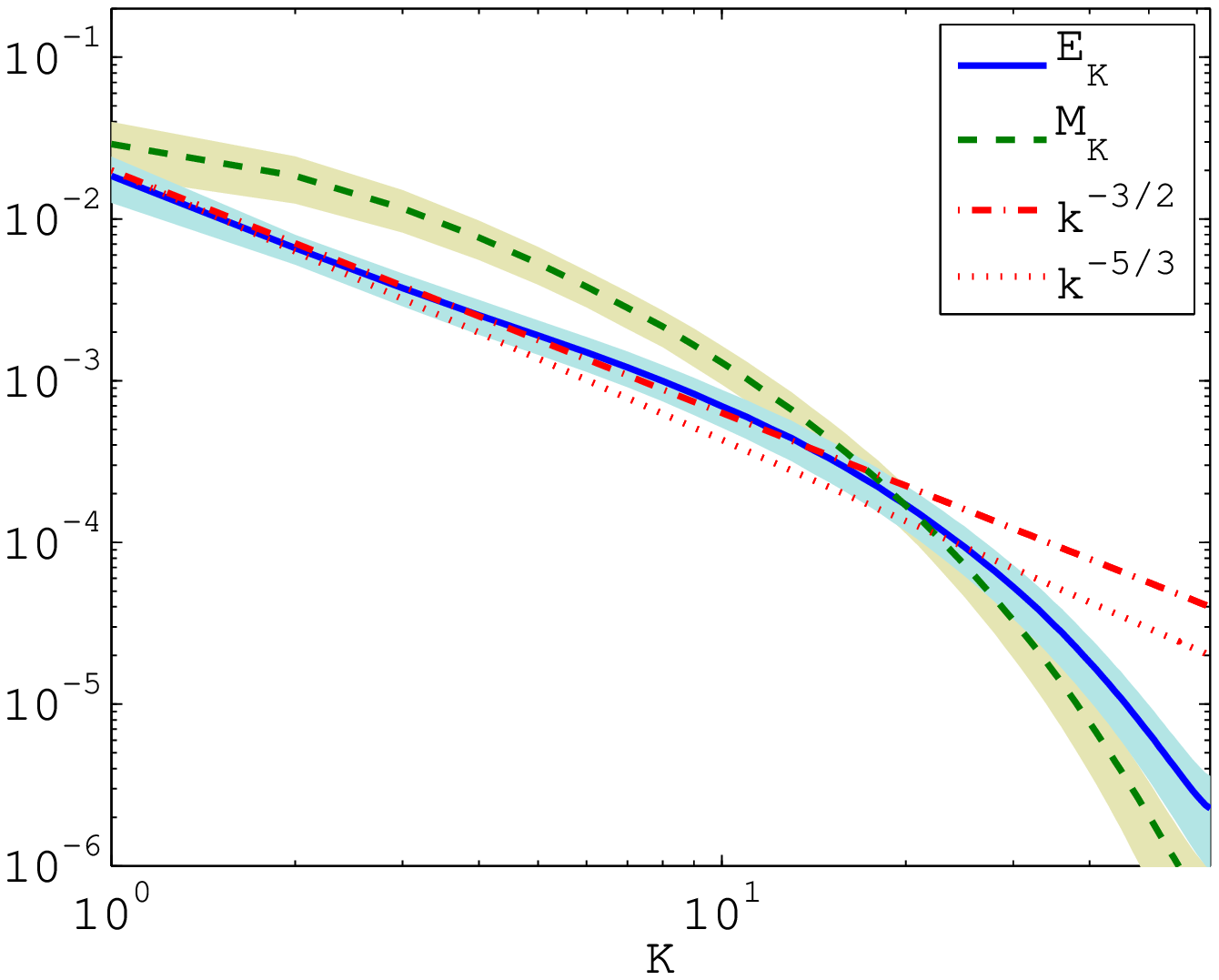}
   \caption{Energy spectrum at $Pm=0.0625$ (left) and $Pm=0.25$ (right). In the $Pm=0.25$ case, a power law spectrum is observed for the kinetic energy corresponding to a $k^{-3/2}$ spectrum.}
              \label{Espectrum}%
\end{figure*}

 \begin{figure*}[t]
   \centering
   \includegraphics[width=0.45\linewidth]{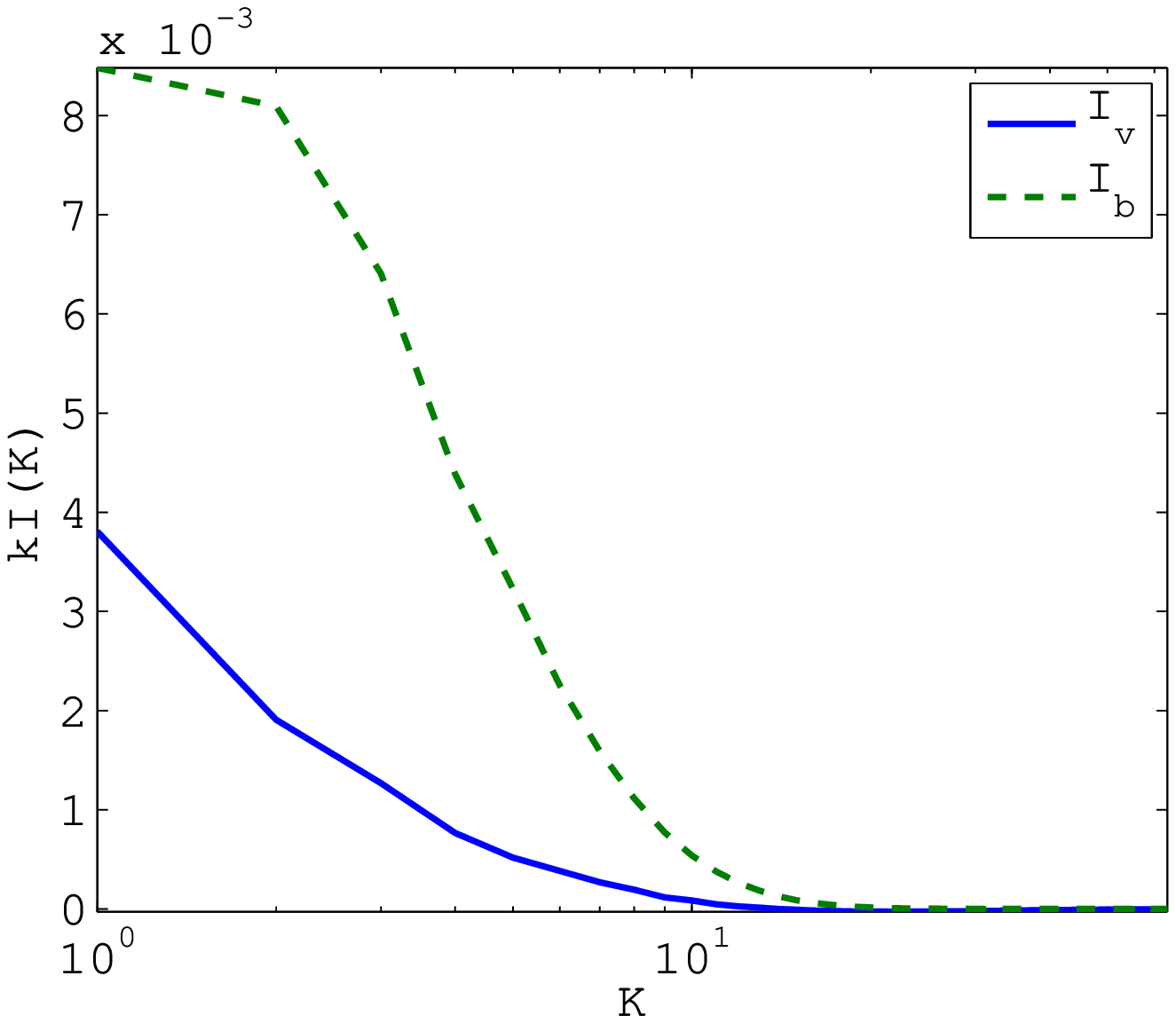}
   \includegraphics[width=0.45\linewidth]{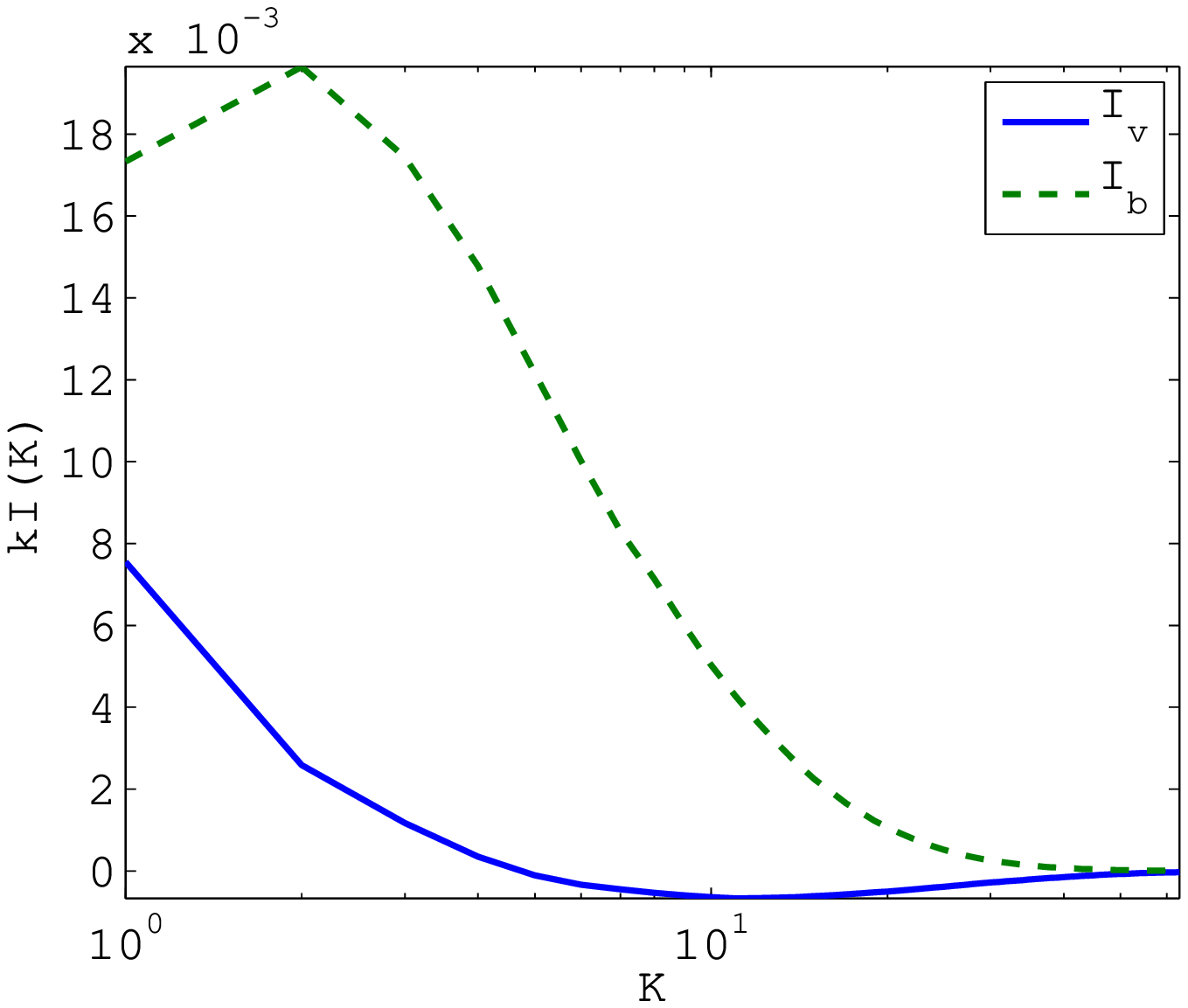}
   \caption{Energy Injection spectrum at $Pm=0.0625$ (left) and $Pm=0.25$ (right). Although the injection is significantly reduced at small $Pm$, shape of the spectrum is similar and dominated by the largest scale.}
              \label{Ispectrum}%
\end{figure*}

\subsection{Numerical method}

Equation (\ref{motion}) and (\ref{induction}) are solved using the Snoopy code. Snoopy is a 3D spectral (Fourier) method based on the shearing wave decomposition (\ref{shwaveX})---(\ref{shwaveZ}). The Fourier transforms are evaluated using the FFTW 3 library, with both MPI and OpenMP parallelisation techniques. Nonlinear terms are computed using a pseudo-spectral algorithm \citep{C88} and antialiasing is enforced using the ``3/2'' rule. Time integration is performed by a third order, low storage Runge-Kutta scheme for non-linear terms, whereas an implicit scheme is used for viscous and resistive terms. This spectral scheme uses a periodic remap algorithm in order to continuously follow the smallest wave number of the system in the sheared frame $|\bm{k}(t)|<k_\mathrm{max}$ (see \citealt{UR04} appendix C for a complete description of the periodic remap algorithm). Note that the periodic remap method used in this code is different from the continuous remap method used by \cite{Li07}. Our main motivation to implement a periodic remap is the possibility of using the 3/2 antialiasing rule and power of 2 grid sizes\footnote{If the number of point in one direction is a multiple of 2, waves at the Nyquist frequency do not have any properly defined phase. This is not a problem for classical spectral methods or for the periodic remap method since the Nyquist frequency is either in the dissipation range or in the antialiasing dump zone. However, when using a continuous remap method, the Nyquist frequency waves are remapped to large scale waves in physical space, which might lead to non-physical behaviours.} for which Fourier transform and parallelisation methods are more efficient. This code or its variant has been used in several context, including the MRI \citep{LL10} and the subcritical baroclinic instability \citep{LP10}. It is available for download on the author's website. 

\section{Results\label{results}}
\subsection{Simulations parameters and averaging procedure}

The spectra and transfers presented in this section are all derived from two simulations of an MRI saturated state. These runs corresponds to the $Pm=0.25$ and $Pm=0.0625$ high resolution runs discussed in \cite{LL10}.
Both runs have a resolution\footnote{The quoted resolution now includes the aliasing domain. This convention differs from the one adopted in our previous papers, where only the "useful" domain was accounted for when quoting resolutions.} $N_x\times N_y\times N_z=768\times 384\times 192$ with a box aspect ratio $L_x\times L_y\times L_z=4\times 4\times 1$. We impose a mean vertical field in the box with $\beta=10^3$ and $\Lambda_\nu=30$ ($Re=2\times 10^4$).  Each simulation is integrated for 50 orbits starting from random noise and the spectra are averaged from the last 40 orbits to remove any influence of the initial conditions. The two simulations considered in this section only differ by their ohmic resistivity, the $Pm=0.25$ run having $\Lambda_\eta=7.5$ ($Rm=5000$) whereas the $Pm=0.0625$ run has $\Lambda_\eta=1.87$ ($Rm=1250$). 

The statistical average of any quantity $X$ of interest $\langle X \rangle_{stat}$ should in principle be computed on different realizations but are evaluated in practice as usual via an ergodic hypothesis:

\begin{equation}
\langle X \rangle_{stat} = \lim_{T\rightarrow ±\infty} \frac{1}{T}\int_0^T\ X(t)dt \simeq \frac{1}{T}\int_0^T\ X(t)dt\simeq \sum_i X(T_i),
\end{equation}

\noindent where $0 \le T_i \le T$ are a sufficiently large number of instants of flow snapshots.

The spectra are averaged in the spherical shells introduced in (\ref{Xshell}). The shells are defined so that $K_n=2\pi n/L_z$ and $\delta K=2\pi/L_z$. This means that some power is present in the shell $K=0$, as it contains large scale horizontal waves with no vertical structure. The shell-integrated spectra and transfers obtained by this procedure are then averaged in time over 40 instantaneous snapshots (1 snapshot per orbit). For simplicity, we have renormalized the wavevectors $K$ so that $K'=K/2\pi$ on all the plots in this section. Note that shells $K>32$ are incomplete in the $y$ direction  since the resolution per scale height is smaller in that direction. This is not a problem since these shells are in the dissipative range and high $k_y$ modes are weaker than the equivalent high $k_x/k_z$ modes due to the anisotropy of MRI turbulence (see section \ref{sec:spectra}). Note that, with the procedure we have used, one can reconstruct the box averaged quantities by summing the spectra over the integers K.

The shear transfer terms $F_{s,v}$ and $F_{s,b}$ are computed in a special way. Indeed, one cannot compute $F_s$ for a given shell and snapshot time numerically because of the $\delta$ functions. Instead, we introduce a shell-averaged flux:

\begin{equation}
F_s^c(K_0,t)=\frac{1}{\delta K}\int_{K_0-\delta K/2}^{K_0+\delta K/2}dK\,\mathcal{F}_{s}(K,t).
\end{equation}

As $\langle \mathcal{F}_{s} \rangle_{stat}$ depend only on $K$ (the turbulence is statistically stationary) and varies little with $K$ on scales of the order of  $\delta K$, one has $\langle \mathcal{F}_{s} \rangle_{stat} \simeq $$\langle F_s^c \rangle_{stat}$. One can therefore use $F_s^c$ in the averaging procedure described above to estimate $\langle \mathcal{F}_{s} \rangle_{stat}$.

The numerical flux we obtain is then averaged over time following the procedure described above.

\subsection{\label{sec:spectra}Spectra and energy injection}

 \begin{figure*}[t!]
   \centering
   \includegraphics[width=0.38\linewidth]{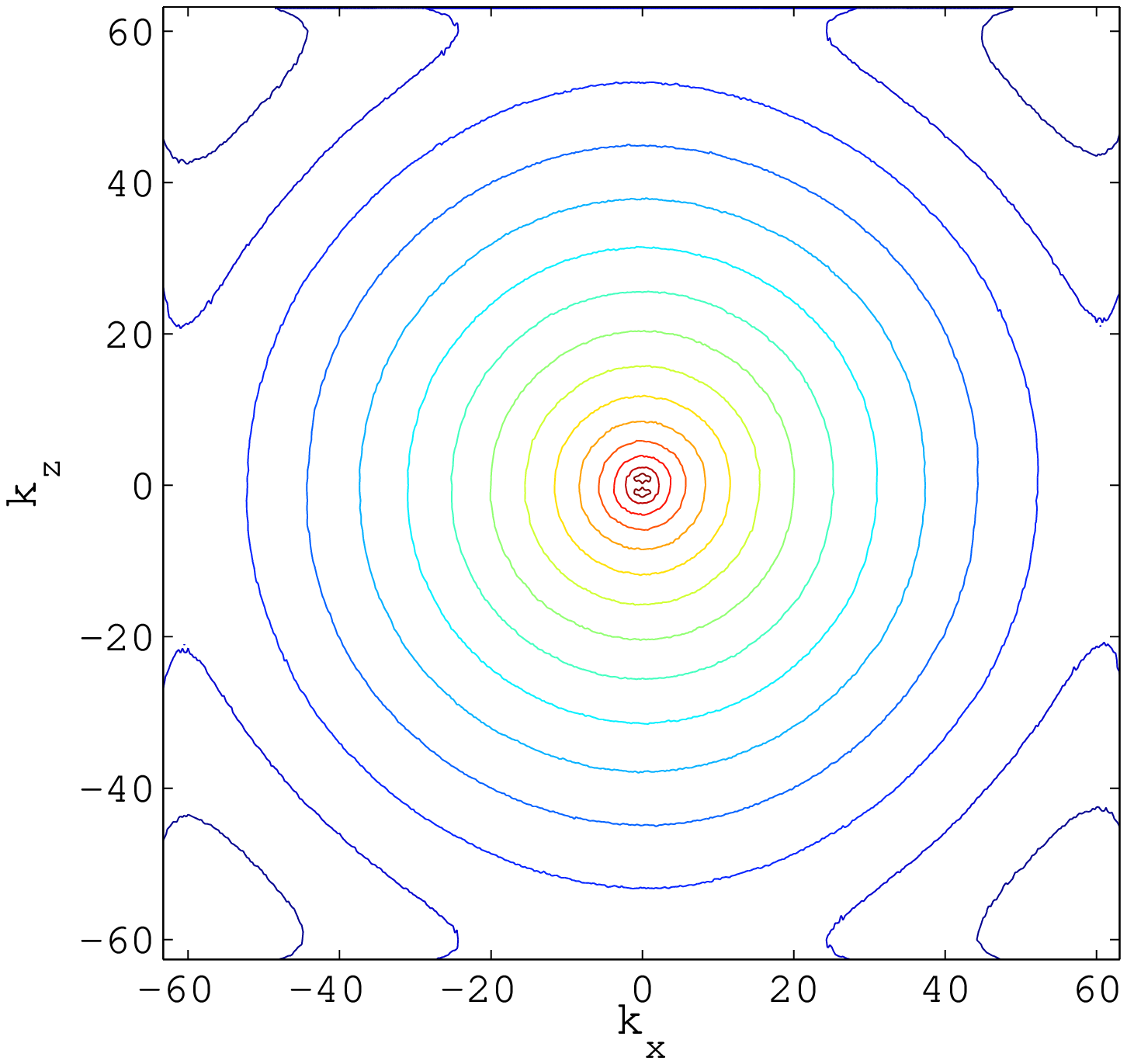}
   \includegraphics[width=0.38\linewidth]{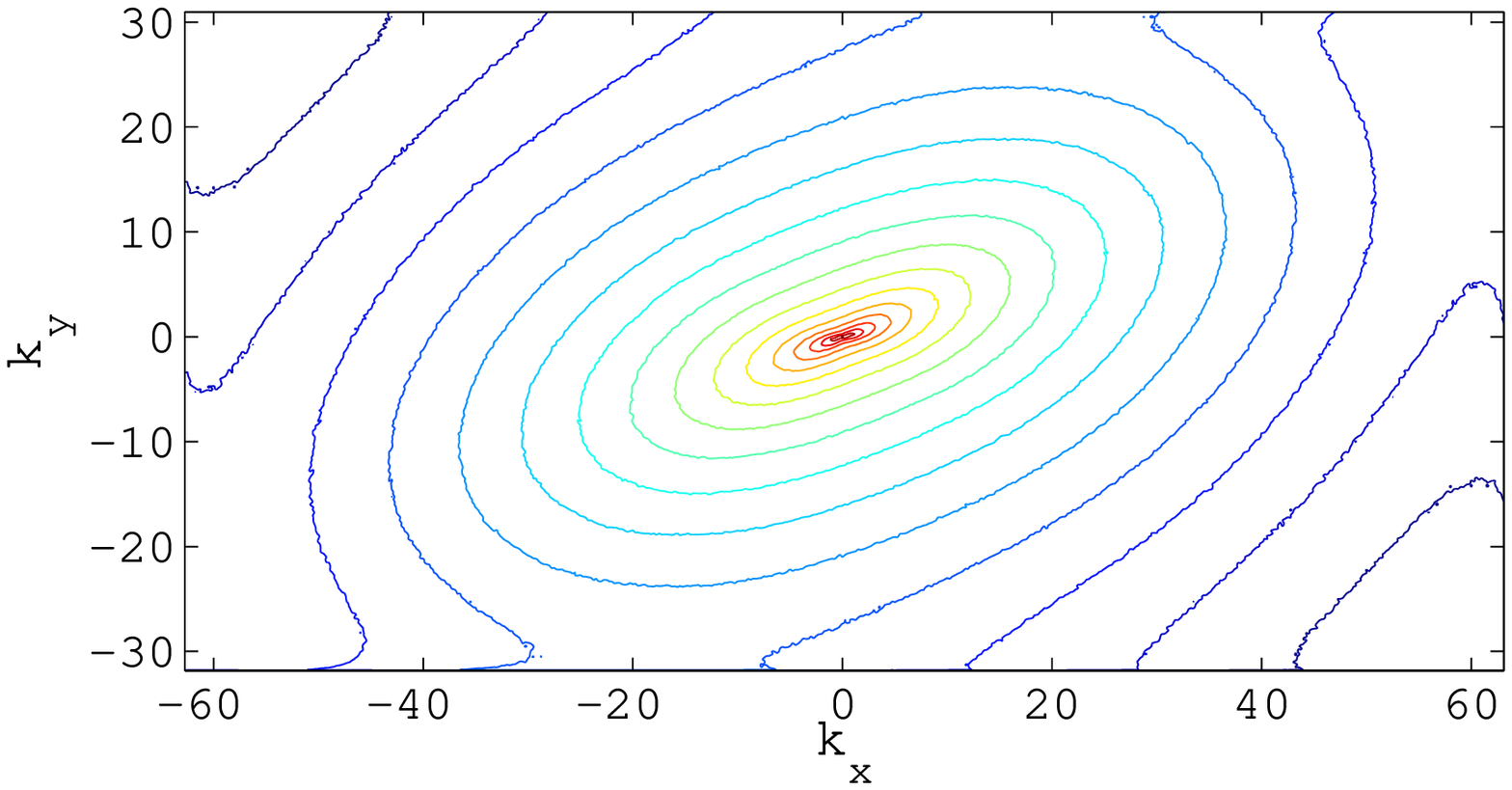}
   \includegraphics[width=0.22\linewidth]{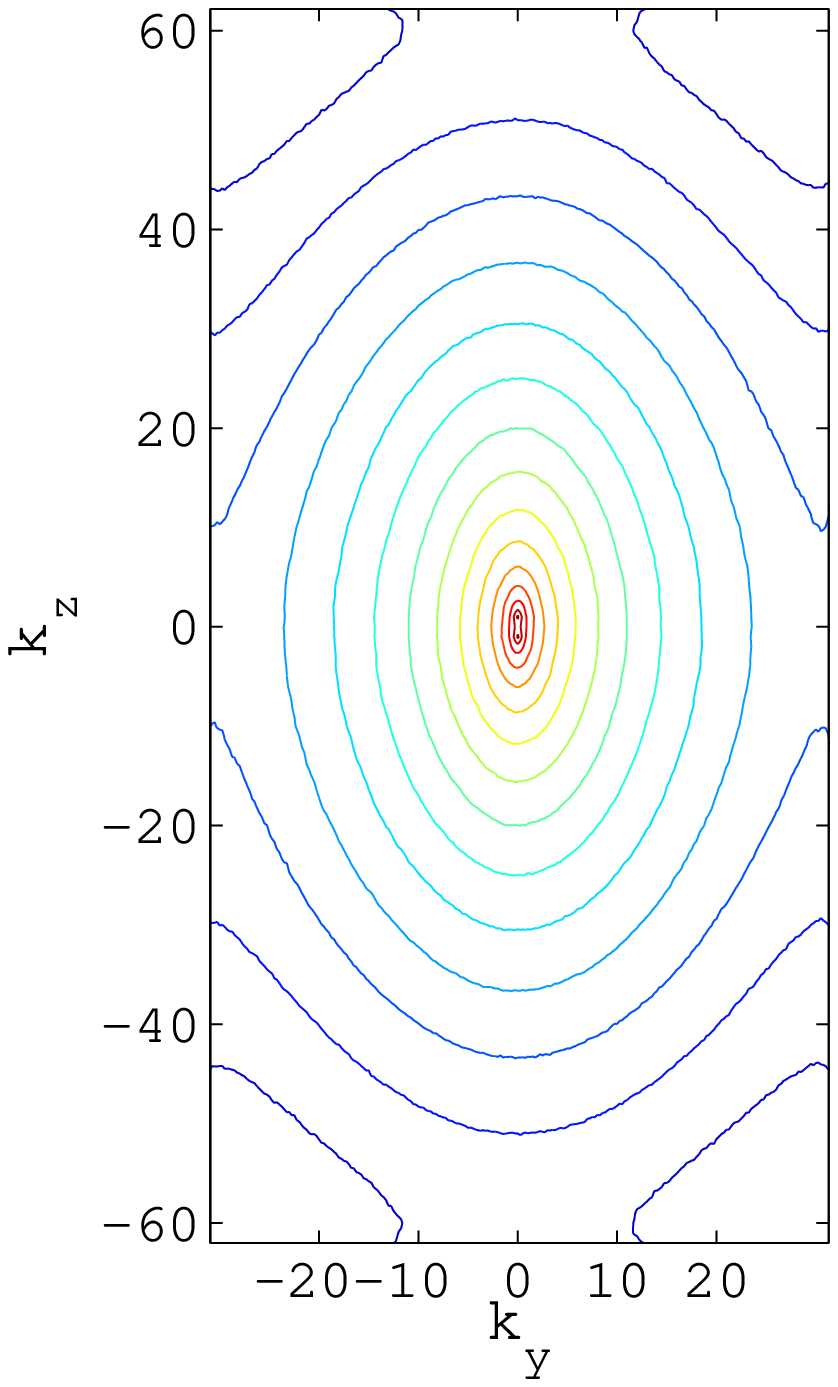}
   \caption{Bidimensional magnetic energy spectra at $Pm=0.25$. Left: $(k_x, k_z)$ spectrum, averaged in $k_y$. Centre: $(k_x, k_y)$ spectrum, averaged in $k_z$. Right: $(k_y, k_z)$ spectrum averaged in $k_x$. Each contour correspond to a factor 10 in magnetic energy.}
              \label{spectrum2D}%
\end{figure*}

We first present the energy spectra on Fig.~\ref{Espectrum} for $Pm=0.0625$ and $Pm=0.25$. The standard deviation, measured from 40 instantaneous snapshots, is shown as a shaded region on these spectra. This dispersion is due to temporal fluctuations of the turbulence intensity. The most obvious feature observed in these spectra is the presence of a $k^{-3/2}$ spectrum for the kinetic energy; the traditional Kolmogorov scaling $k^{-5/3}$ appears to be excluded in the $Pm=0.25$ run, but it cannot be strictly excluded in the $Pm=0.0625$ run. A $k^{-3/2}$ power-law was also found in zero net flux MRI turbulence \citep{F10}, although the spectrum shape differs, our spectra being exempt of any ``bump'' at intermediate scale. As our runs do not yet resolve the inertial range of the turbulent cascade, these apparent spectral shapes require some comment. The presence of a $k^{-3/2}$ spectrum is usually related to the theoretical argument of \cite{I63} and \cite{K65} (or shortly IK).  However MRI turbulence is not strongly magnetized, and therefore falls outside the domain of validity of the IK phenomenology. Moreover, the magnetic field spectrum does not follow any well-defined power law, as expected from the wide and overlapping injection (see below) and dissipation spectra, indicating that the spectrum we get is \emph{not} an inertial spectrum. We are therefore forced to conclude that although the kinetic spectrum looks like an IK {or Kolmogorov} spectrum, it is not described by the IK or Kolmogorov phenomenologies, nor by recent extensions \citep{B05}.

Changing the magnetic Prandtl number does not change the power-law index for the kinetic energy. We note however 2 major effects: the overall spectra amplitudes are reduced and the dissipation scales move to larger scale as one reduces $Pm$. These two effects are expected since it is known that smaller $Pm$ turbulence is associated with a smaller transport efficiency and therefore a weaker injection of energy in the cascade. This effect is confirmed by the injection spectra (Fig.~\ref{Ispectrum}) which are significantly reduced at smaller $Pm$. 

We note that the energy injection peaks at the largest scale of the box, although injection still exist at $k\sim10$. Therefore, although a power-law spectrum is found for $2<k<10$, this spectrum cannot be described as an ``inertial range'' since energy is still injected at these intermediate scales.

We present on Fig.~\ref{spectrum2D} bidimensional spectra of magnetic energy for $Pm=0.25$ (kinetic spectra are not shown as they share essentially the same properties). These spectra were obtained averaging 3D energy spectra over 40 orbits and taking the average in the $k_x$, $k_y$, or $k_z$ directions. We first note a strong anisotropy in the $(k_x, k_y)$ plane which indicates that trailing shearing waves ($k_xk_y>0$) have more energy than leading shearing waves ($k_xk_y<0$). As we will see below, this results in non zero shear transfer terms. 

Looking at the aspect ratio of the energy contours, we see that turbulence is slightly less anisotropic at large $k$ than at small $k$ (the contours are less ``elongated'' at large $k$), although complete isotropy is not yet reached in this simulation. Let us however remark that the spectral truncation (due to the finite resolution) tends to deform the contours at large $k$, which might accelerate the return to isotropy. One should therefore perform higher resolution runs (or at least double $N_y$) in order to confirm this return to isotropy. In principle, one would expect a return to isotropy at small scales if the nonlinear transfer terms dominate all the other terms (injection, body forces) at large enough $k$. However this is not always the case (e.g. in the presence of a strong mean magnetic field).

The $(k_x,k_z)$ spectrum shows that turbulence is essentially isotropic at large $k$ in that plane. For $k\sim 1$, we find a slight anisotropy where modes with $k_z\neq 0$ are favoured. This is probably a result of large scale MRI unstable modes which all have $k_z\ne 0$ in the presence of a mean vertical field. Note that this anisotropy disappears very quickly as one moves to larger $k$.

\subsection{Magnetic helicity and cross helicity}

\begin{figure*}[t]
   \centering
   \includegraphics[width=0.45\linewidth]{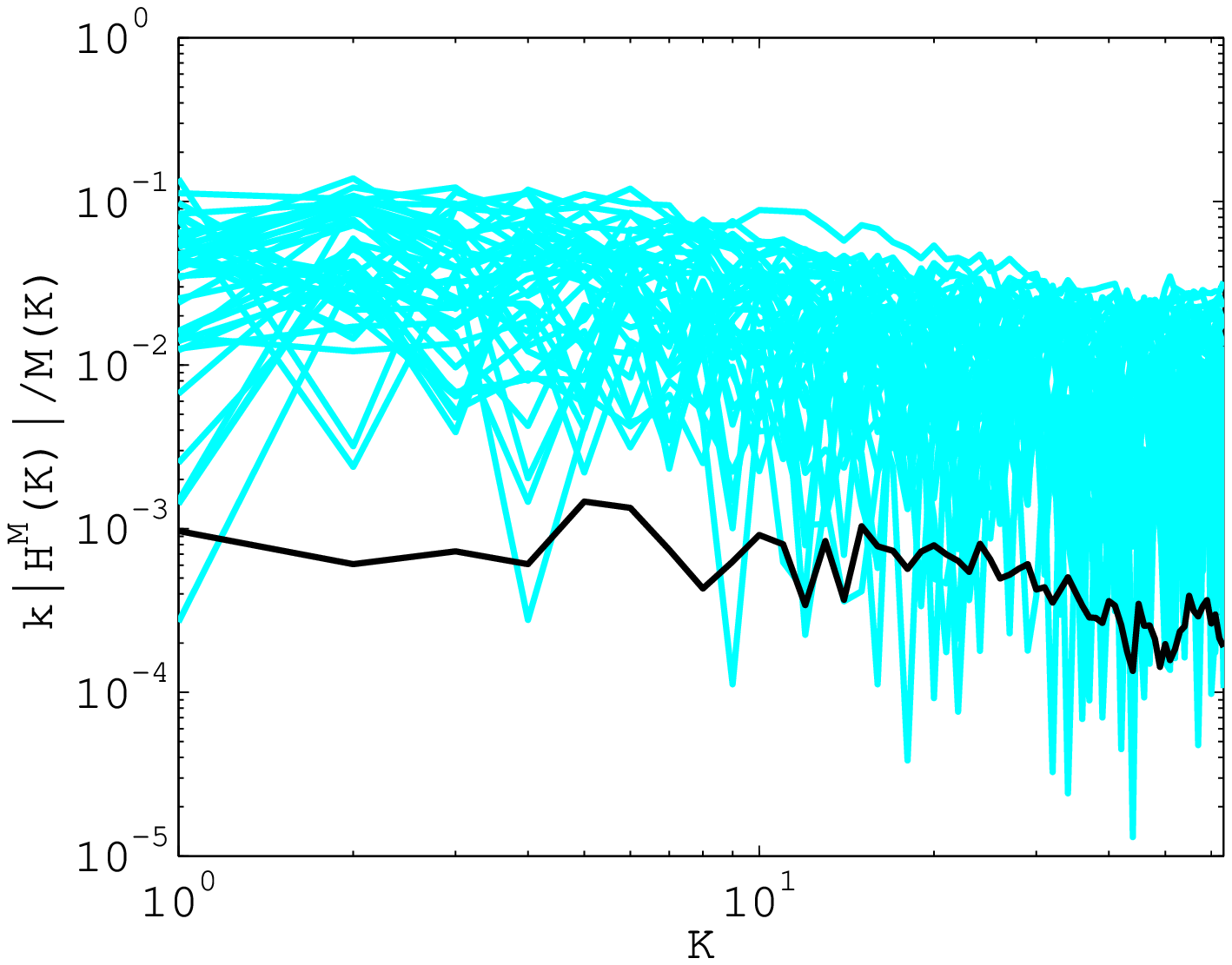}
   \includegraphics[width=0.45\linewidth]{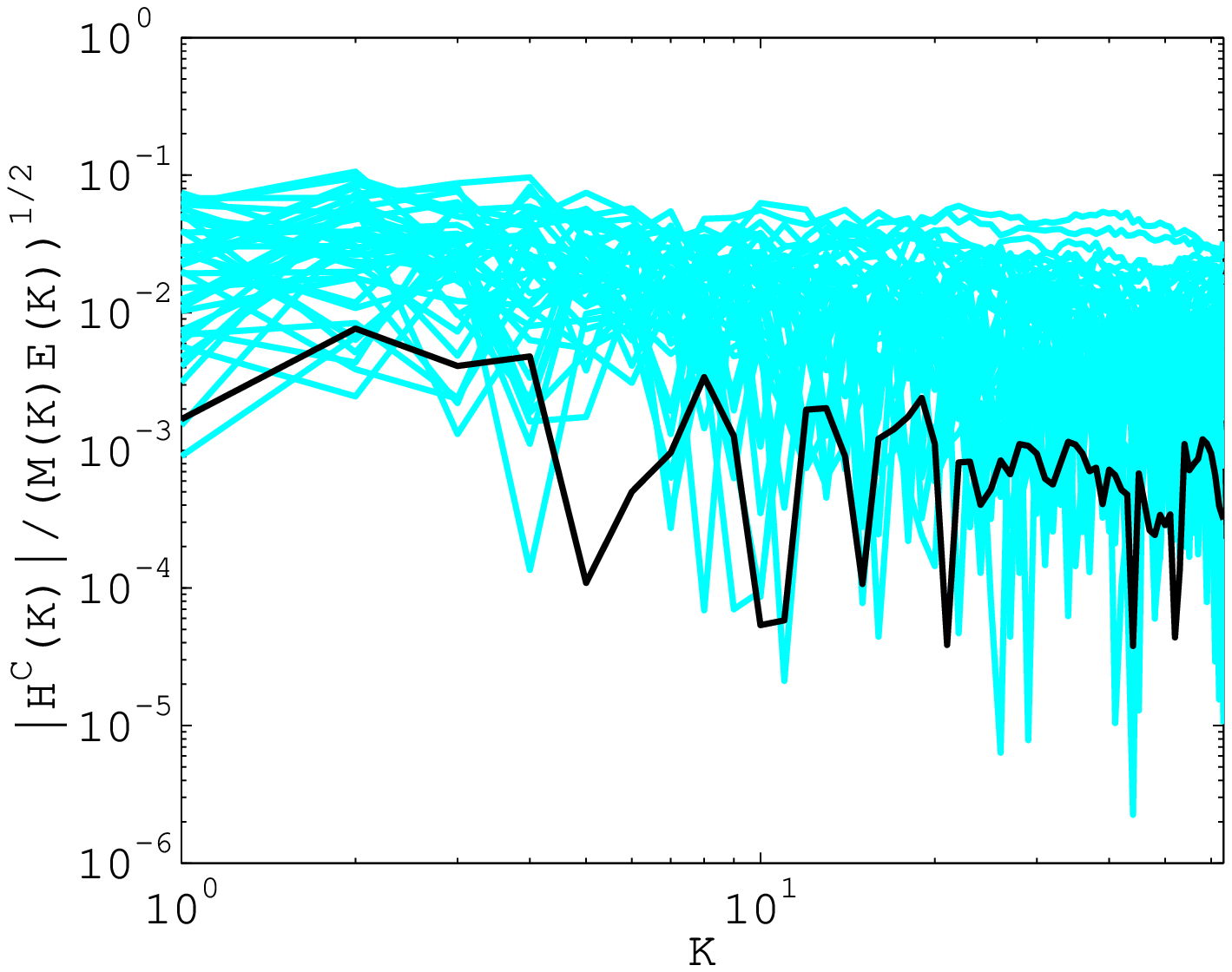}
   \caption{Average relative magnetic helicity (left) and cross helicity (right) spectra in the $Pm=0.25$ case (black lines). Instantaneous spectra are represented in light blue. Note that the absolute value of relative helicities is plotted here, as helicities sign is constantly changing.}
              \label{HelSpectrum}%
\end{figure*}
\begin{figure*}[t!]
   \centering
   \includegraphics[width=0.45\linewidth]{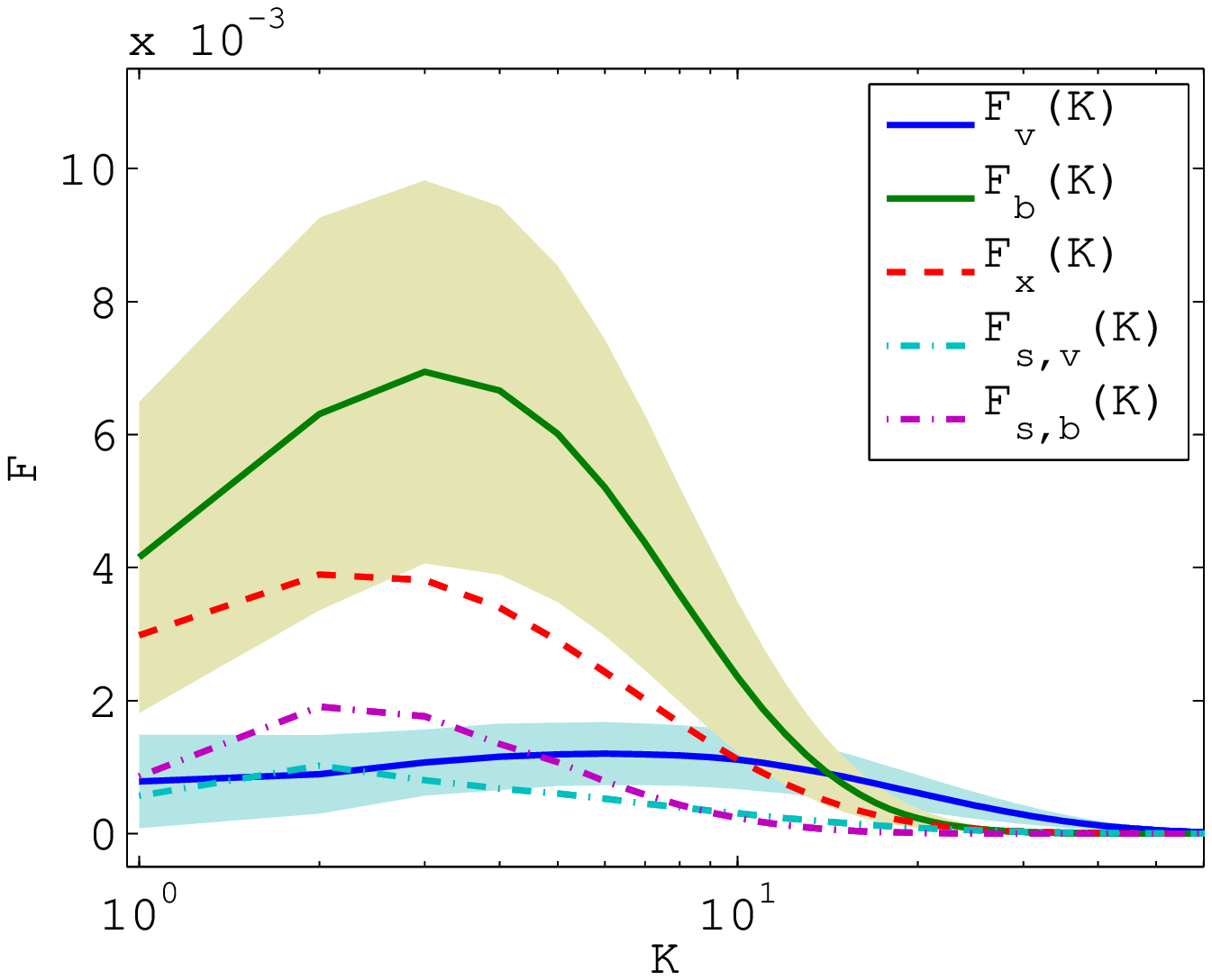}
   \includegraphics[width=0.45\linewidth]{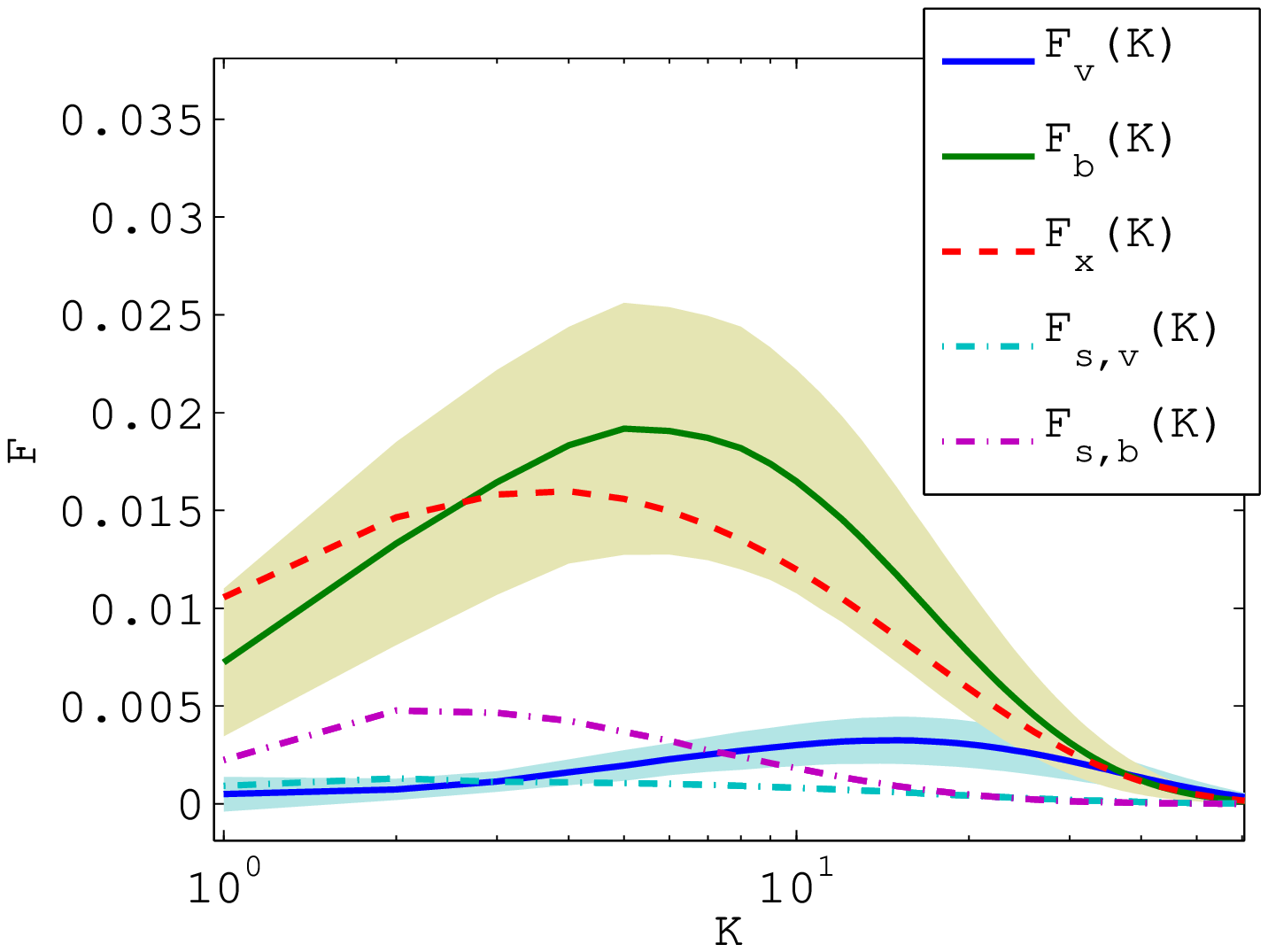}
   \caption{Energy fluxes as computed from equation (\ref{Eq:Flux1})-(\ref{Eq:Flux2}) at $Pm=0.0625$ (left) and $Pm=0.25$ (right). Energy fluxes are always direct (from large to small scale) and dominated by magnetic and exchange fluxes.}
              \label{FluxSpectrum}%
\end{figure*}

The presence of kinetic and/or magnetic helicity in MHD turbulence is often invoked to explain large scale dynamo action. Indeed, it is known that an inverse cascade of magnetic helicity can appear in fully developed helical MHD turbulence \citep{FP75}, potentially leading to a buildup of magnetic energy at large scale. Some kinematic effects often used in mean field dynamos, like the $\alpha$ effect \citep{M78}, also lead to the generation to large scale helical fields \citep{BS05}. Magnetic helicity has therefore been suggested as a possible driving mechanism (or at least a tracer) of disc dynamos \citep{B10}. Moreover, several authors have tried to link magnetic helicity conservation and magnetic helicity flux to the saturation properties of the MRI \citep{V09,KK10}.

In this work, we define the magnetic helicity $\mathcal{H^M}=\langle\bm{A}\cdot(\bm{B}-\langle \bm{B}\rangle)\rangle$, where $\bm{A}$ is the vector potential of the fluctuation (this expression is gauge-invariant in the shearing box);
$\langle \cdot \rangle$ denotes a volume average. We show on Fig.~\ref{HelSpectrum} (left) the spectrum of relative helicity $K |H_K|/2M_K$ for the $Pm=0.25$ run. As it can be seen, the relative helicity is less than 1\% for all scales of these simulations. Moreover, this quantity is strongly fluctuating and its sign is not well defined\footnote{Note that the absolute value of the relative magnetic helicity is plotted on Fig.~\ref{HelSpectrum}}.

These results tend to indicate that magnetic helicity is dynamically unimportant in the unstratified simulations presented here and that MRI saturation is not related to a magnetic quenching effect due to magnetic helicity accumulated at large scale. This was to be expected in the first place as unstratified shearing boxes (both with and without mean field) are mirror symmetric. Note however that this picture might change when stratification is included.

An other quantity of interest which might play a role in the MHD turbulence cascade is the cross helicity $H^C=\langle \bm{v\cdot b}\rangle$ (see e.g. \citealt{PB10}). When cross helicity is non zero, the energy of Alfv\'en waves traveling along and against the mean field are not equal. For this reason, turbulence with cross helicity is often called \emph{imbalanced} turbulence. Locally imbalanced turbulence is often observed in strong MHD turbulence, where the guide field is weaker than the turbulent fluctuations. To check whether non stratified MRI turbulence was imbalanced, we have computed cross helicity spectra of our simulations [Fig.~\ref{HelSpectrum} (right)]. As for the magnetic helicity, we find that the relative cross helicity is small ($<10^{-1}$) and highly fluctuating at all scales. This tends to indicate that MRI turbulence is not imbalanced in our setup. As for magnetic helicity, this result was to be expected because of the mirror symmetry properties of the non stratified shearing box. The absence of any significant cross helicity also shows that energy spectra in Els\"asser variables $z^\pm=v\pm b$ are equal and proportional to the kinetic plus magnetic energy spectrum.

\subsection{Energy fluxes}

The energy fluxes (\ref{Eq:Flux1})-(\ref{Eq:Flux2}) allow one to check the average direction of the energy flux in spectral space. To explain the dependence of the turbulent transport on $Pm$, several authors (e.g. \citealt{LL07,FPLH07}) have suggested that an inverse cascade driven by resistive and viscous scales might be at work. Since magnetic helicity is irrelevant to this problem, only the kinetic, magnetic, exchange and shear fluxes are important for the non stratified shearing box and should be checked for an inverse cascade.

We present the energy fluxes at $Pm=0.25$ and $Pm=0.0625$ on Fig.~\ref{FluxSpectrum}. Standard deviations are shown for kinetic and magnetic fluxes as shaded regions. These deviations are computed following the procedure described in section \ref{sec:spectra}.
We always find positive fluxes, meaning that the non linear transfers are \emph{forward} or direct (from large to small scales) on average. However, at larger scale, the standard deviation may allow for an inverse cascade of kinetic energy in the $Pm=0.25$ run. This indicates that, although the kinetic cascade is direct on (time) average over most of the spectrum, inverse cascades can sometimes be observed on the largest scales. This inverse cascade of kinetic energy could be an explanation to the large scale hydrodynamic structures which are observed in several MRI turbulence simulations, such as vortices \citep{FN05} and zonal flows \citep{JYK09}.

We also observe that the energy cascade is dominated by the magnetic and exchange fluxes down to the resistive scale, the kinetic flux and shear fluxes being almost negligible. Note also that the shear fluxes are always positive. This is due to the anisotropy of MRI turbulence in which shearing waves have statistically a larger amplitude than leading waves (see section \ref{sec:spectra}).

\begin{figure}[t]
   \centering
   \includegraphics[width=0.95\linewidth]{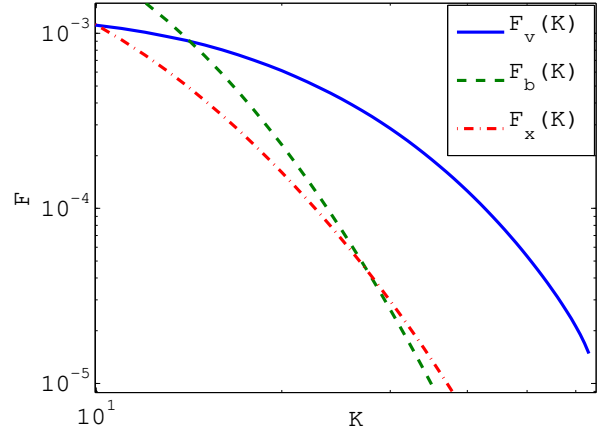}
   \caption{Zoom on the energy fluxes in the dissipative range for $Pm=0.0625$ (log-log representation with $K>10$).}
              \label{FluxSpectrumZoom}%
\end{figure}

In the $Pm=0.0625$ case, the kinetic flux becomes dominant at subresistive scale ($K\gtrsim 10$), indicating that the cascade becomes essentially hydrodynamic below the resistive scale, as expected for small $Pm$ MHD turbulence (see also Fig.~\ref{FluxSpectrumZoom}). Moreover, the kinetic flux dominates the kinetic shear transfer term at least for the larger $k$, which indicates that, as far as the non linear transfers are concerned, the cascade is close to isotropic at small scales, as noted in section \ref{sec:spectra}. Finally, note that none of the fluxes reach a plateau at intermediate scales, which would be expected in the presence of an inertial range. This indicates that the $k^{-3/2}$ kinetic spectrum found in Fig.~\ref{Espectrum} is not properly speaking an inertial spectrum.

\begin{figure*}[t!]
   \centering
   \includegraphics[width=0.3\linewidth]{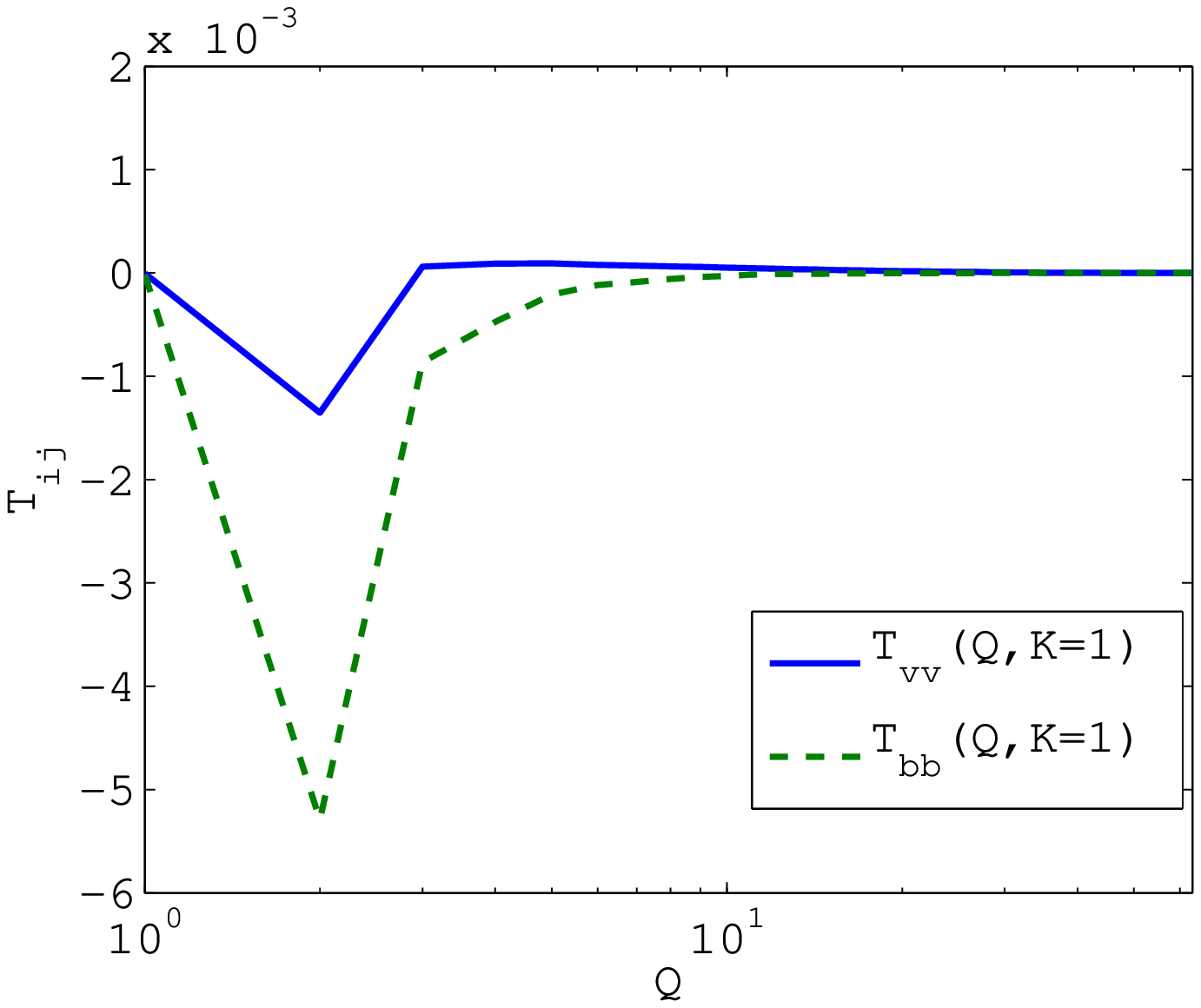}
   \includegraphics[width=0.3\linewidth]{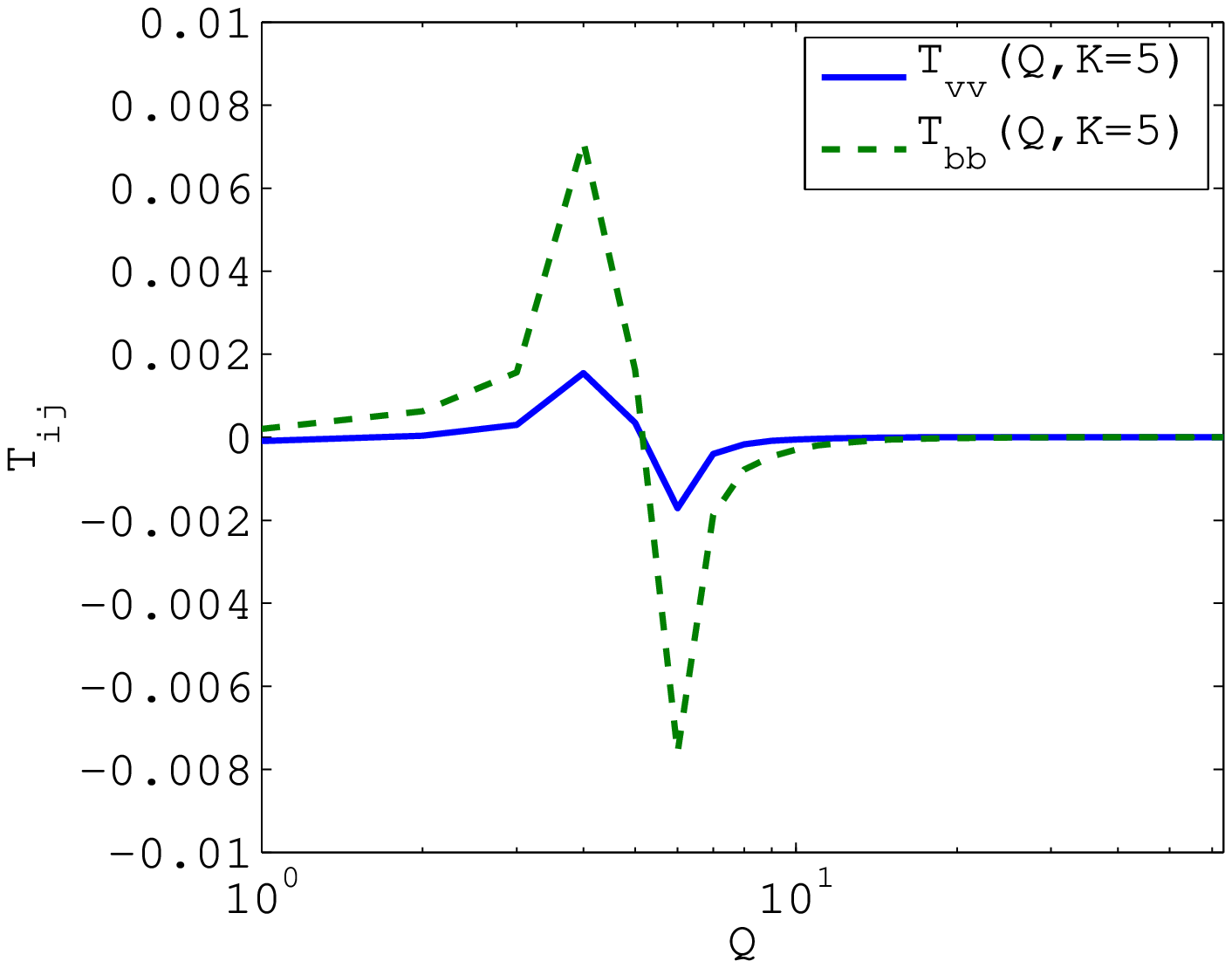}
   \includegraphics[width=0.3\linewidth]{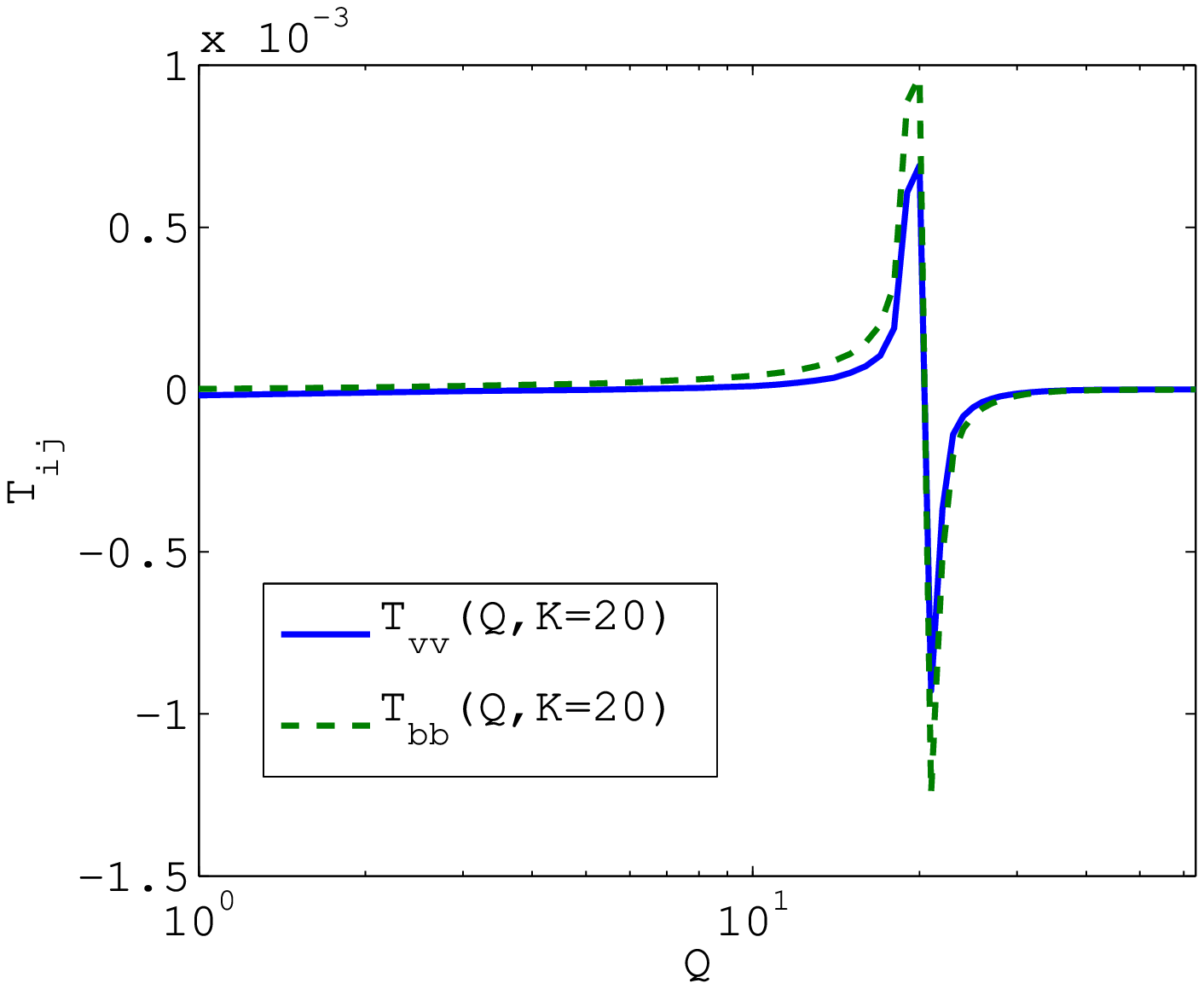}
   \caption{Transfert function $T_{vv}(Q,K)$ and $T_{bb}(Q,K)$ in the $Pm=0.25$ run for $K=1; 5; 20$. These transfers are \emph{local} in Fourier space (see text).}
              \label{TransfertVV}%
\end{figure*}
\begin{figure*}[t!]
   \centering
   \includegraphics[width=0.3\linewidth]{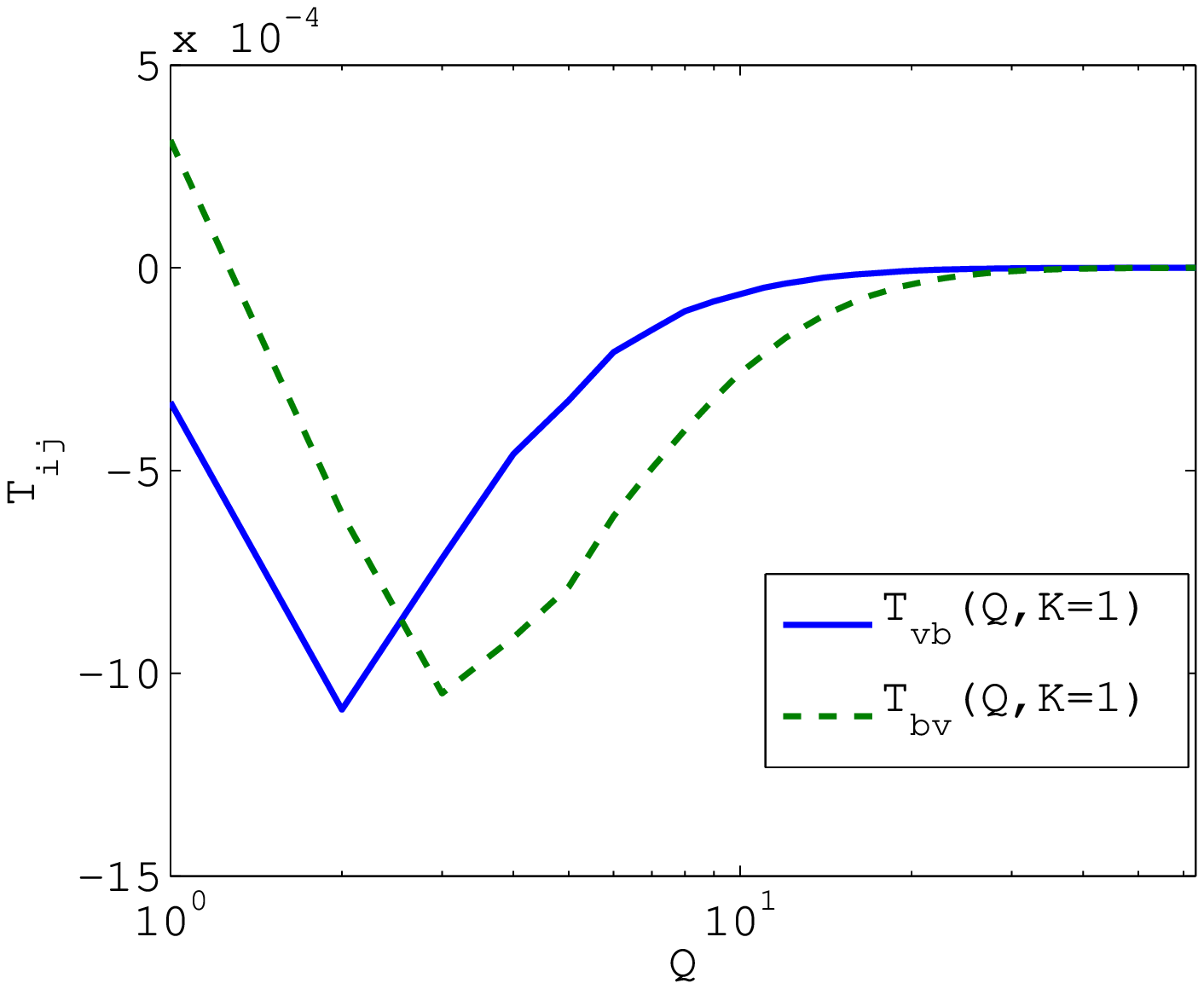}
   \includegraphics[width=0.3\linewidth]{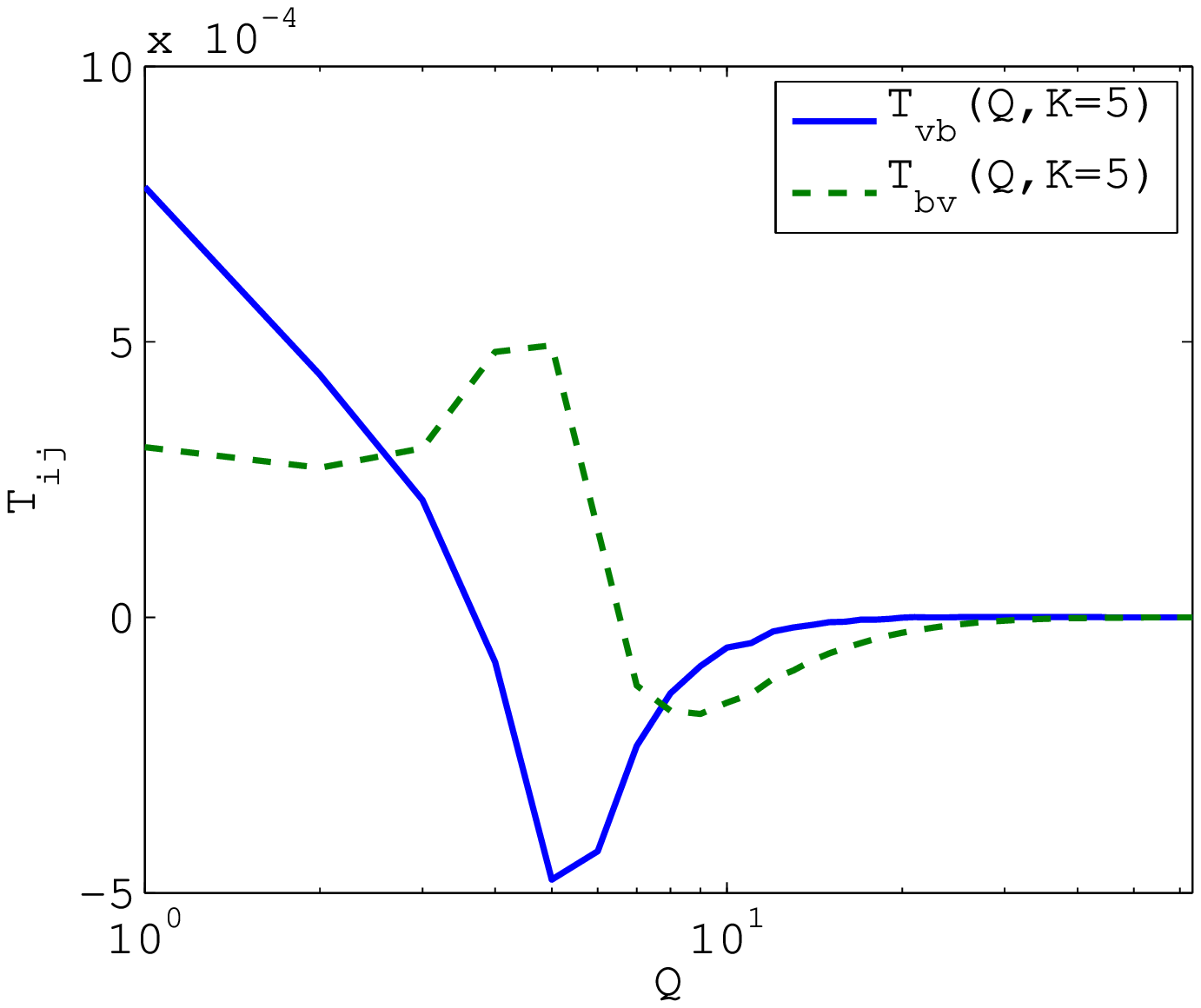}
   \includegraphics[width=0.3\linewidth]{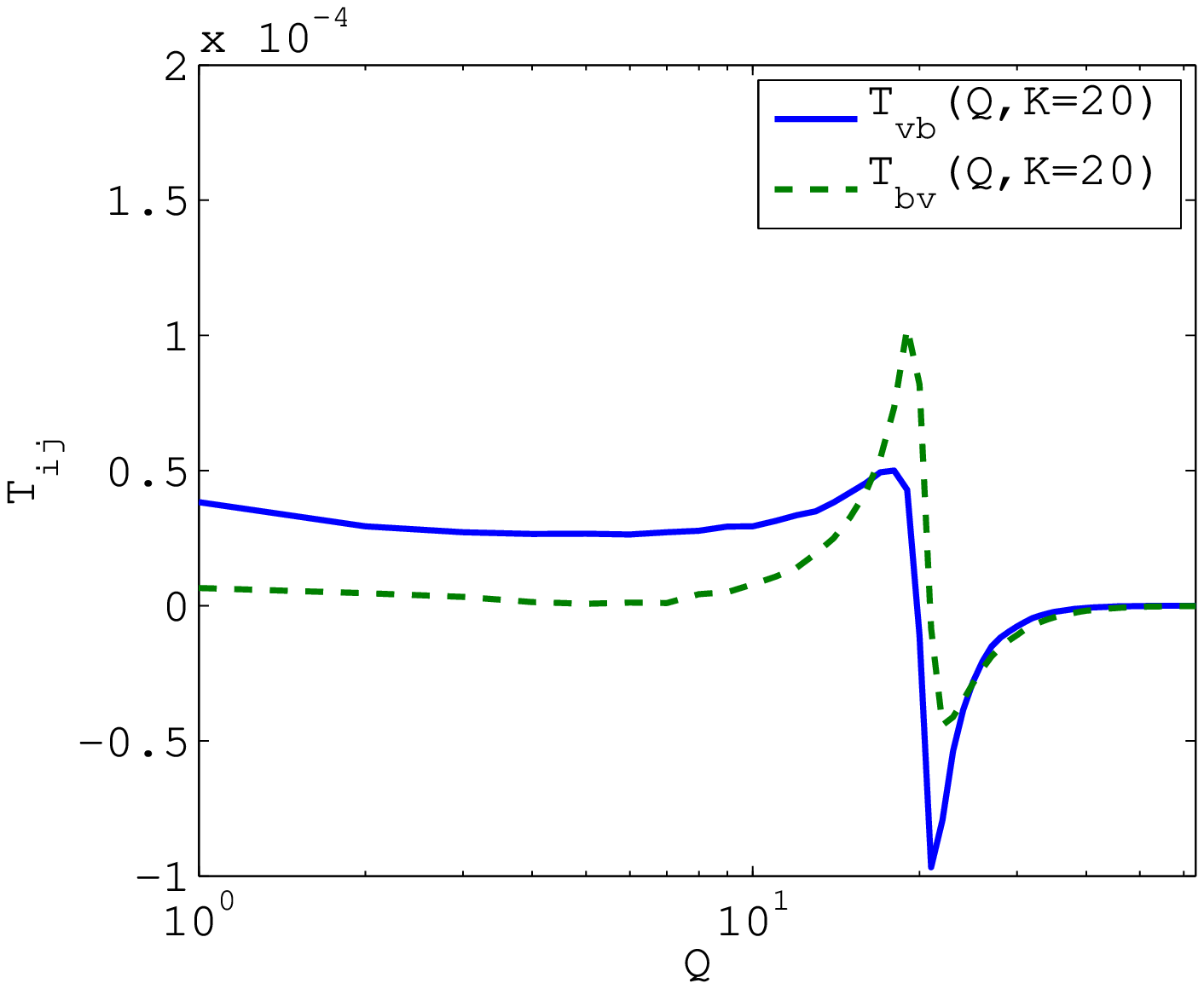}
   \caption{Transfert function $T_{vb}(Q,K)$ and $T_{bv}(Q,K)$ in the $Pm=0.25$ run for $K=1; 5; 20$. These transfers are \emph{non-local} in Fourier space (see text).}
              \label{TransfertVB}%
\end{figure*}

\subsection{Energy transfer locality}

To test the locality of the energy transfer in spectral space, we have plotted the transfer functions in the $Pm=0.25$ case for several values of $K$: at the injection scale ($K=1$), in the intermediate range ($K=5$) 
and in the resistive range ($K=20$). We first plot the kinetic to kinetic and magnetic to magnetic transfers on Fig.~\ref{TransfertVV}. The transfers $T_{vv}$ and $T_{bb}$ obtained at all scales perfectly illustrate \emph{local} energy exchanges. Energy is taken from wavenumbers slightly smaller than $K$ and is transferred to wavenumbers slightly larger than $K$, except (not surprisingly) for $K=1$. As expected from the energy flux, we also find that the cascade is direct, with energy going from small to large wavenumbers, Finally, note that the $T_{vv}$ transfers are always much smaller than the $T_{bb}$ transfers above the dissipation range, illustrating the magnetically dominated energy transfer described above.

We next plot the exchange transfers $T_{vb}$ and $T_{bv}$ on Fig.~\ref{TransfertVB}. We note in this case that the scales involved in each transfer are much broader. In particular, $T_{vb}$ measured at the resistive scale ($K=20$) has contribution coming from all scales, including the largest injection scales. This effect can also be seen in $T_{bv}(Q,K=1)$, which exhibit a very long tail toward large $k$, down to the resistive scale. Comparing directly these transfers to $T_{vv}$ and $T_{bb}$ show that these terms are \emph{highly non local}. In turns, this indicates that the exchange flux computed in the previous section is non local. The results in the $Pm=0.0625$ case are not shown here as they are very similar to the $Pm=0.25$ case.

We note that despite the non locality of the energy exchanges, the overall cascade direction is still forward, confirming our previous interpretation regarding the exchange flux. We also remark that the shear transfer terms are local \emph{by definition} as they transfer energy to neighbouring shells.

\section{Discussion}

In this paper, we have described some properties of the turbulent cascade found in incompressible MRI turbulence. We have shown that compared to isotropic MHD turbulence, the presence of a mean shear led to several new transfer terms and introduced a source of anisotropy. We have computed the effect of each non linear term and found that all the terms contribute to a direct cascade of energy (from large to small scales) but some terms involved non local transfers in Fourier space. This non locality is due to the Lorentz force and to the magnetic stretching term of the induction equation (combined here in the exchange transfer term). We have also shown that magnetic helicity, although non zero, was totally negligible and should not play any role in the behaviour of MRI turbulence. 

The presence of non local transfer terms in the MRI turbulent cascade is the most important finding of this work. It indicates that in principle, the large scales -- responsible for the transport -- can \emph{directly} interact with the small dissipative ones through non linear terms. This direct interaction could of course explain the correlation observed between $Pm$ and the turbulent transport of angular momentum $\alpha$ \citep{LL10}. However, it should be pointed out that nonlocal transfers were already found in isotropic MHD turbulence \citep[e.g.][]{AMP07}. Therefore, MRI turbulence has nothing special regarding the nature of these non linear transfers. 

Although some nonlinear terms are found to directly connect injection and dissipation scales in current simulations, one might wonder if this would be true in a more realistic setup where the injection and dissipation scales are separated by a wide range of scales (typically $10^{10}$). In other words, what is the maximum scale separation these terms can connect?
A partial answer to this question is given by \cite{AE10}. In order to describe their result, let us define the structure functions:
\begin{equation}
\delta v_{\bm{l},p} =\langle \vert\bm{v}(\bm{x}+\bm{l})- \bm{v}(\bm{x})\vert^p\rangle\nonumber
\end{equation}
In the inertial range the structure function depends only on $\vert\bm{l}\vert$ and $\delta v_{l,p}\propto l^{\zeta_p}$, where $\zeta_p$ is the structure function index of order $p$. It is then possible to derive an upper bound to the nonlinear transfer terms thanks to the H\"older inequality. Applying this procedure to the non local transfer $T_{ub}$, \cite{AE10} found
\begin{equation}
\label{Ubound}
|T_{ub}(Q,K)| < \mathrm{(const)} Q^{1-\zeta^u_3/3}K^{-2\zeta^b_3/3}
\end{equation}
where $Q$ and $K$ are dyadic (octave) wavenumbers and $K>Q/2$. Similar terms can be obtained for $T_{bu}$ and $K<Q/2$. If one assumes Iroshnikov-Kraichnan theory, one has $\zeta^u_3=\zeta^b_3=3/4$. On the contrary, considering Goldreich-Sridhar (GS) phenomenology, which should be valid for MRI turbulence, one gets $\zeta^u_{3\parallel}=\zeta^b_{3\parallel}=3/2$ and $\zeta^u_{3\perp}=\zeta^b_{3\perp}=1$. In all these cases, (\ref{Ubound}) indicates that the non-locality of these transfer terms cannot extend over several decades, with a typical scaling $T_{ub}(Q,K)\sim \epsilon (K/Q)^{-2/3}$ for GS turbulence ($\epsilon$ being the usual turbulence energy injection rate). 

We therefore conclude that the non locality in Fourier space is somewhat relative. Although $T_{ub}$ and $T_{bu}$ are non local compared to $T_{uu}$ and $T_{bb}$, these terms should be local when one considers transfers over several decades. Unfortunately, separating the injection scale from the dissipative scales by several decades is numerically difficult. It is even harder for MRI turbulence since the injection term is rather broad in spectral space compared to forced turbulence. Assuming the injection and dissipation scales both spread over one decade in Fourier space, one typically needs $20,000^3$ simulations to get a 2 decades inertial range in which non local transfers are significantly reduced. This kind of resolution is for the moment out of reach of the best computational facilities. 

Nevertheless, we can conjecture that if the $Pm-\alpha$ correlation is effectively due to the non local transfers, then it should vanish when the injection and dissipation scales are well separated as it is the case in some accretion discs. Although this conclusion looks rather reassuring for the relevance of today simulations regarding small scale dissipation, it does not tell us what the asymptotic value of $\alpha$ is in this limit, nor how MRI turbulence behaves when the scale separation is \emph{not} achieved, a situation which probably occurs in the inner regions of protoplanetary discs where $\Lambda_\eta$ is not very large.

\appendix
\section{Shearing waves approach to the shear transfer term\label{shearTransfer}}
\begin{figure}[t!]
   \centering
   \includegraphics[width=0.9\linewidth]{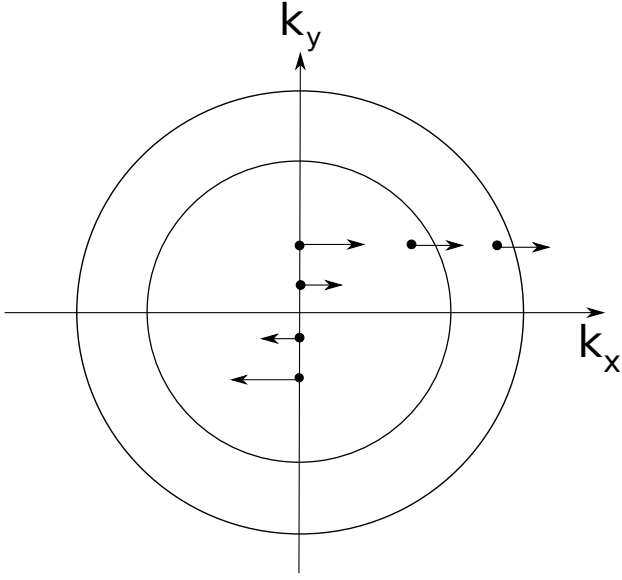}
   \caption{Evolution of shearing waves in the presence of a fixed shell in Fourier space. Some waves can either enter or exit the shell as time evolves (see text).}
              \label{FigShell}%
\end{figure}

The shell-filter decomposition can be properly defined using a projector operator $\Pi$ on the field $F$ in the sheared frame:
\begin{equation}
[\Pi_{K_j}(F)](x',t)=\sum_{\bm{k}'\in\Sigma_j(t)} F_{\bm{k'}}(t) \exp\left(i {\bm{k'}}\cdot {\bm{x'}}\right)\nonumber
\end{equation}
where $\Sigma_j$ is the shell containing all the shearing waves with a norm between $K_j-\delta K/2$ and $K_j+\delta K/2$: 
\begin{equation}
\Sigma_j(t) = \Big \{ K_j-\delta K/2 < |\bm{k}'+q\Omega k_y't\bm{e}_x| \le K_j+\delta K/2\Big \}
\end{equation} 
Our notation indicates that the projected $F$ is function of space and time. Note also that $[\Pi_{K_j}(F)]$ is real for real fields $F$.

As it can be seen, the waves included in $\Sigma_j(t)$ change with time. This is to be expected as shearing waves should in principle enter and exit shells as the one defined above (see Fig.~\ref{FigShell}). As a result, the projector operator $\Pi_{K_j}$ has an explicit time dependence which leads to non trivial transfer effects. The above projector operator can be written using Heaviside function $\Theta$:
\begin{eqnarray}
\nonumber [\Pi_{K_j}(F)](x',t)&=&\sum_{\bm{k}'}  \Theta\Big(|k(t)|-K_j+\delta K/2\Big)\\
\nonumber &&\times \quad\Theta\Big(-|k(t)|+K_j+\delta K/2\Big)\\
\label{ProjectHeaviside}& &\times\quad F_{\bm{k'}}(t) \exp\left(i {\bm{k'}}\cdot {\bm{x'}}\right).
\end{eqnarray}
where we have defined $\bm{k}(t)$ as a function of $\bm{k}'$ as in (\ref{shwaveX})---(\ref{shwaveZ}).

One next defines the energy within a shell $E_{K_j}^F=\langle \Pi_{K_j}^2(F)\rangle/2$ where $\langle\cdot\rangle$ denotes a volume average. The energy equation then reads
\begin{eqnarray}
\nonumber \frac{dE_{K_j}^F}{dt}&=&\Big\langle \Pi_{K_j}(F)\frac{\partial\Pi_{K_j}(F)}{\partial t}\Big\rangle\\
\label{Eappendix}&=&\big\langle \Pi_{K_j}(F)\Pi_{K_j}(\partial_tF)\big\rangle+\big\langle \Pi_{K_j}(F)\big[\partial_t\Pi_{K_j}\big](F)\big\rangle
\end{eqnarray}
The first term on the right handside leads to the terms obtained in isotropic turbulence \citep{AMP07} and introduced in Eqs.~(\ref{EqEk}) and (\ref{EqEm}) along with the injection terms $I_{v,K}$ and $I_{b,K}$. The second one however is due to
shearing waves entering and exiting the shells. 
Using (\ref{ProjectHeaviside}), it is possible to obtain an exact (though singular) expression of the operator time derivative:
\begin{eqnarray}
\label{ShearTerm}\nonumber \partial_t\big[\Pi_{K_j}\big](F)&=&\sum_{k'}     \frac{q\Omega k_y k_x(t)}{|k(t)|}\\
\nonumber & &\times \Big[\delta\Big(|k(t)|-K_j+\delta K/2\Big)- \delta\Big(-|k(t)|+K_j+\delta K/2\Big)\Big]\\
 & &\times\,\, F_{\bm{k'}}(t) \exp\left(i {\bm{k'}}\cdot {\bm{x'}}\right).
\end{eqnarray}
This expression can be interpreted easily. As an example, let us consider waves with $k_yk_x(t)>0$. Then, the first $\delta$ function represents waves entering the shell, the second delta represents waves leaving the same shell, and the factor in front of the delta functions quantifies the ``flux'' of waves going through a shell boundary. This interpretation is similar to the phenomenological picture one can have of waves traveling through a fixed shell in the unsheared Fourier space (Fig.~\ref{FigShell}). 

We then deduce from (\ref{Eappendix})
\begin{eqnarray}
 \label{Efinal} \frac{dE_{K_j}^X}{dt}&=&\langle \Pi_{K_j}(F)\Pi_{K_j}(\partial_tF)\rangle\\
\nonumber & &+ \sum_{k'}     \frac{q\Omega k_y k_x(t)}{|k(t)|}\frac{F_{\bm{k'}}^* F_{\bm{k'}}}{2}\\
\nonumber & &\times \Big[\delta\Big(|k(t)|-K_j+\delta K/2\Big)- \delta\Big(-|k(t)|+K_j+\delta K/2\Big)\Big].
\end{eqnarray}
where we have used the property\footnote{This can be shown differentiating the relation $\Theta^2(x)=\Theta(x)$} $\delta(x)\theta(x)=\delta(x)/2$. As expected, this expression shows two contributions to the energy evolution inside a shell: a volume contribution and a surface contribution equal to the energy of the waves entering and leaving the shell. Introducing the equation of motions in sheared space (\ref{motion-shear})---(\ref{induction-shear}) in the above relation leads to the energy equations (\ref{EqEk})---(\ref{EqEm}).

Note finally that using the relation
\begin{equation}
\partial_t|k(t)|=\frac{q\Omega k_y k_x(t)}{|k(t)|},
\end{equation}
one can write (\ref{Efinal}) as
\begin{equation}
\frac{dE_{K_j}^X}{dt}=\langle \Pi_{K_j}(F)\Pi_{K_j}(\partial_tF)\rangle+ \sum_{k'}  \frac{F_{\bm{k'}}^* F_{\bm{k'}}}{2}\delta(t-t_{k'}) \epsilon_{k'}
\end{equation}
where $t_{k'}$ is the instant when the wave $k'$ enters or exits the shell $K_j$ and $\epsilon_{k'}= \pm1$ for an entering/exiting wave. This somewhat simpler expression has the same interpretation as (\ref{Efinal}).

\section{Unsheared Fourier transform approach to the shear transfer term\label{shearTransfer2}}

It is possible to understand the origin of the shear transfer term (\ref{ShearTerm}) starting from the equations of motion in unsheared coordinates (\ref{motion})---(\ref{induction}) with an appropriate use of continuous Fourier transforms. Shear-periodic functions are not absolutely integrable, but this difficulty can be circumvented because their continuous Fourier transform is well-defined as a distribution. To demonstrate this point, let us consider a 2D infinite medium in which a field $F$ obeys the model equation
\begin{equation}
\label{AppModel}\partial_t F(\bm{x},t)-q\Omega x\partial_y F(\bm{x},t)=0.
\end{equation}
Let us introduce the \emph{unsheared} Fourier transform:
\begin{eqnarray}
\nonumber \tilde{F}(\bm{k},t)&=&\frac{1}{(2\pi)^3}\iiint d\bm{x}\,F(\bm{x},t)\exp(-i\bm{k}\cdot\bm{x})\\
\label{AppIFT}F(\bm{x},t)&=&\iiint d\bm{k}\,\tilde{F}(\bm{k},t)\exp(i\bm{k}\cdot\bm{x}).
\end{eqnarray}

If $F$ obeys the shearing sheet boundary conditions, the solution of the equation is a Fourier series $F_p$ of the form (\ref{SpectrumDef}) with Fourier coefficients $F_{\bm{k'}}$. Its Fourier transform in the unsheared spectral space is then:
\begin{equation}\label{FTF}
\nonumber \tilde{F}_p(\bm{k},t)=\sum_{k'} \delta(\bm{k}-\bm{k}_0(t))F_{\bm{k'}}
\end{equation}
where $\bm{k}_0(t) = \bm{k}'+\bm{V}_{k'} t$ and where we have defined the ``Fourier velocity'' $\bm{V}_k = q\Omega k_y \bm{e_x}$, evaluated in $\bm{k}'$. By construction $\tilde{F}_p(\bm{k},t)$ is solution of:
\begin{equation}
\label{AppModelFourier}
\partial_t\tilde{F}(\bm{k})+q\Omega k_y \partial_{k_x}\tilde{F}(\bm{k})=0,
\end{equation}
which is the Fourier transform of our model equation\footnote{The Fourier transform of (\ref{AppModel}) gives (\ref{AppModelFourier}) except for an extra term that vanishes once the Fourier series expression of $F$ is used.}. The left handside of this equation can be interpreted as a comoving derivative of $\tilde{F}(k,t)$ in Fourier space with the Fourier velocity $\bm{V}_k$. This equation tells us that the amplitude of the waves is constant when one moves at velocity $\bm{V}_k$ in Fourier space consistently with the form of the solution $\tilde{F}_p$. It can also be interpreted as constant amplitude shearing waves, as expected.

It is then possible to introduce the projector operator, now time-independent as it is defined in unsheared coordinates:
\begin{eqnarray}
\nonumber [\Pi_{K_j}(F)](x,t)&=&\iiint d\bm{k}\,  \Theta\Big(|k|-K_j+\delta K/2\Big)\\Ê
\nonumber & & \times\quad \Theta\Big(-|k| + K_j + \delta K/2\Big)\\
& &\times\quad \tilde{F}(\bm{k},t) \exp\left(i {\bm{k}}\cdot {\bm{x}}\right).
\end{eqnarray}
As previously, the shell energy ($E_{K_j}^F=\langle \Pi_{K_j}^2(F)\rangle/2$) time variation follows from:

\begin{eqnarray}
\nonumber \frac{dE_{K_j}^F}{dt} &=& \big\langle \Pi_{K_j}(F)\partial_t\big[\Pi_{K_j}(F)\big]\big\rangle\\
&=&-\big\langle \Pi_{K_j}(F)\Pi_{K_j}\big(\bm{\nabla}_{\bm{k}}\cdot(\bm{V}_k\tilde{F})\big)\big\rangle
\end{eqnarray}

\noindent where (\ref{AppModelFourier}) has been used in the last equality as well as  $\bm{\nabla}_{\bm{k}}\cdot\bm{V}_k =0$. Because the volume average selects the zero frequency contributions, this leads us back to (\ref{Efinal}) with the help of the relation
\begin{eqnarray}
\nonumber \int dk_x\,\Theta(|k|-C)\partial_{k_x} \tilde{F}(\bm{k}) &=&-\int dk_x\,\frac{k_x}{|\bm{k}|}\delta(|k|-C)\tilde{F}(\bm{k}).
\end{eqnarray}
and with the use of $\delta(x)\theta(x)=\delta(x)/2$ and (\ref{FTF}).

In the spirit of the integral expressions used in this appendix, Eq.~(\ref{Efinal}) can be recast in integral form by introducing the energy density in Fourier space $\mathcal{E}^F(\bm{k},t)$ defined by

\begin{equation}
\mathcal{E}^F(\bm{k},t)=\sum_{k'} \delta(\bm{k}-\bm{k}_0(t))\frac{F_{\bm{k'}}F^*_{\bm{k'}}}{2}.
\end{equation}

With this definition,

\begin{equation}
E_{K_j}^F=\iiint d\bm{k}\,  \Theta(|k|-K^-)\Theta(-|k| + K^+)\ \mathcal{E}^F(\bm{k},t)
\end{equation}

\noindent while
\begin{eqnarray}
\frac{dE_{K_j}^F}{dt} &=& \iiint d\bm{k}\, \bm{V_k\cdot n}\ \mathcal{E}^F(\bm{k},t)\big[\delta(|k|-K^-)-\delta(-|k| + K^+)\big]\nonumber\\
& & \\
&=& -\oiint _{\partial K_j}  d\bm{k}\, \bm{V_k\cdot n}\ \mathcal{E}^F(\bm{k},t)
\end{eqnarray}

In these relations, $K^{\pm}=K\pm\delta K/2$ has been defined; the second integration is performed on the surface of the shell $K_j\equiv \partial K_j$ and $\bm{n}$ is the normal to this surface. The last expression has an explicit flux form.

This approach can be applied to the original shearing sheet MHD equations; the time dependence due to the shear term will produce the shear flux contribution just computed.

\begin{acknowledgements}
 GL acknowledges support by STFC. PYL acknowledges the hospitality of the \textit{Isaac Newton Institute} and of the DAMTP in Cambridge where parts of this research has been conducted. GL thanks S. Fromang for his comments on the initial version of the manuscript. This work was granted access to the HPC resources of IDRIS under allocation x2009042231 made by GENCI (Grand Equipement National de Calcul Intensif).
\end{acknowledgements}

\bibliographystyle{aa}
\bibliography{glesur}

\end{document}